\documentclass[a4paper,11pt]{article}
\pdfoutput=1 

\usepackage{jheppub} 
\usepackage{amssymb}
 
\newcommand*{\FigPath}{./figs/}%
\newcommand*{\BibPath}{.}%

\newcommand{\mup}{\mu}
 
\newcommand{\be}{\begin{equation}}
\newcommand{\ee}{\end{equation}}
\newcommand{\bea}{\begin{eqnarray}}
\newcommand{\eea}{\end{eqnarray}}

\preprint{JLAB-THY-14-1983} 

\title{\boldmath A study on the interplay between perturbative QCD and CSS/TMD formalism in SIDIS processes}

\author[a,b]{M. Boglione,}
\author[b]{J.O. Gonzalez Hernandez,}
\author[a]{S. Melis,}
\author[c]{and A. Prokudin}
 
\affiliation[a]{Dipartimento di Fisica Teorica, Universit\`a di Torino,\\
                Via P.~Giuria 1, I-10125 Torino, Italy}
\affiliation[b]{INFN, Sezione di Torino, Via P.~Giuria 1, I-10125 Torino, Italy}
\affiliation[c]{Jefferson Lab, 12000 Jefferson Avenue, Newport News, VA 23606, USA}

\emailAdd{boglione@to.infn.it}
\emailAdd{joseosvaldo.gonzalez@to.infn.it}
\emailAdd{melis@to.infn.it}
\emailAdd{prokudin@jlab.org}

 \abstract{We study the Semi-Inclusive Deep Inelastic Scattering (SIDIS) cross section as a function of the transverse momentum, $q_T$. 
In order to describe it over a wide region of $q_T$, soft gluon resummation has to be performed. 
Here we will use the original Collins-Soper-Sterman (CSS) formalism; however, the same procedure would hold within 
the improved Transverse Momentum Dependent (TMD) framework.
We study the matching between the region where fixed order perturbative QCD can successfully be applied and the region where 
soft gluon resummation is necessary. We find that the commonly used prescription of matching through the so-called Y-factor cannot 
be applied in the SIDIS kinematical configurations we examine. In particular, the non-perturbative component of the 
resummed cross section turns out to play a crucial role and should not be overlooked even at relatively high energies.
Moreover, the perturbative expansion of the resummed cross section in the matching region is not as reliable as it is usually 
believed and its treatment requires special attention.
}

\begin{document} 

\maketitle

\section{Introduction\label{Intro}}

Calculating the cross section of a  hadronic process at high resolution scale $Q$, where 
a hadron or a lepton pair is experimentally observed over a wide range of transverse momenta $q_T$, 
is a highly non-trivial task. 
While collinear perturbative QCD computations allow us to predict its behaviour in the large $q_T \gtrsim Q$ region, 
diverging contributions of large (double) logarithms arising from the emission of soft and collinear 
gluons need to be resummed in the range of low $q_T$. 
When $q_T\ll Q$, the perturbatively calculated $q_T$ distribution receives large logarithmic contributions, 
proportional to  $(1/q_T^2)\ln(Q^2/q_T^2)$, at every power of $\alpha_s$. Moreover, beyond leading power, 
double logarithms $(1/q_T^2)\ln^2(Q^2/q_T^2)$ are generated, for every power of 
$\alpha_s$, by soft and collinear gluon emissions. Thus, at any order $\alpha_s^n$, the distribution will 
have  logarithmic contributions which become larger and larger as $q_T$ decreases. 
Here $\alpha_s$ cannot be used as the effective expansion parameter of the perturbative series; instead, in this region,  
a perturbative expansion in terms of logarithms is performed, and this perturbative series is then resummed 
into the so-called Sudakov exponential form factor.

This can be achieved by applying a soft gluon resummation scheme like, for instance, the Collins-Soper-Sterman (CSS) 
scheme~\cite{Collins:1984kg}, which  was originally formulated and extensively tested for Drell-Yan (DY) process, 
$h_1 h_2 \rightarrow \ell^+\ell^- X$~\cite{Collins:1984kg,Balazs:1997xd,Qiu:2000hf,Landry:2002ix,Konychev:2005iy}. 
In the case of Semi-Inclusive Deep Inelastic Scattering (SIDIS) process, $\ell N \rightarrow \ell h X$, resummation was studied 
in Refs.~\cite{Nadolsky:1999kb,Koike:2006fn,Su:2014wpa}.

A successful resummation scheme should take care of matching the fixed order hadronic cross section, 
computed in perturbative QCD at large $q_T$, with the so-called resummed cross section,  
valid at low $q_T \ll Q$, where large logarithms are properly treated.
This matching should happen, roughly, at $q_T\sim Q$ where logarithms are small~\cite{Collins:1984kg}, and is very often realized 
through a procedure based on separating the cross section into two parts: one which is regular at small 
$q_T$ (i.e. less singular than $1/q_T^2$) called the Y-term, and one resummed part, called the W-term. 
While the W-term contains the whole essence of resummation itself, the regular Y-term should ensure a 
continuous and smooth matching of the cross section 
over the entire $q_T$ range.

The perturbative resummed series does not converge at extremely low values of $q_T$, where we expect the transverse 
momentum to be ``intrinsic'' rather than generated by gluon radiation.
For the full description of the cross section, one should therefore be able to incorporate in the resummation 
scheme its non-perturbative behaviour. The non-pertubative part of the cross section is subject to 
phenomenological prescriptions and needs to be modeled; however this should, in principle, affect the hadronic 
cross section only in the range where $q_T \to 0$. As a matter of fact we will show that, for low energy SIDIS 
processes (like in COMPASS and HERMES experiments), where $q_T \sim \Lambda_{\rm QCD}$ and $Q$ is 
small (of the order of a few GeV's), the modeled non-perturbative contributions dominate over the entire range of 
measured $q_T$'s. 

Although in this paper we use the CSS resummation scheme, our considerations apply equally well to the TMD 
formalism~\cite{Collins:2011zzd,Aybat:2011zv}. In fact, the cross sections calculated in these two schemes become 
substantially equivalent in phenomenological applications (differing only at higher orders in $\alpha_s$) 
provided one fixes the auxiliary scales $\zeta_F$ and $\zeta_D$ so that: $\zeta_F = \zeta_D= Q^2$~\cite{Aybat:2011zv}.
The correspondence of the two formalisms will be shown explicitly in Appendix~\ref{sec:TMD}.

The paper is organized as follows. In Section~\ref{resummation} we will briefly outline the main steps of resummation 
in a SIDIS process, in the context of the CSS scheme. 
In Section~\ref{sec:match} we will describe some specific matching procedures, discuss the delicate interplay 
between the perturbative and non-perturbative parts of 
the hadronic cross section and give numerical examples, exploring different kinematical 
configurations of SIDIS experiments. Our conclusions will be drawn in Section~\ref{Conclusions}.

\section{Resummation in Semi-Inclusive Deep Inelastic Scattering\label{resummation}}

For unpolarized  SIDIS processes, $\ell N \rightarrow \ell h X$, the following CSS 
expression~\cite{Nadolsky:1999kb,Koike:2006fn} holds 
\begin{align}
\frac{d\sigma^{total}}{d x\, d y \, d z \, d{q}_{T}^2} &= \pi \sigma_0^{D\!I\!S} \!\!\int\frac{d^2
\boldsymbol{b}_T e^{i \boldsymbol{q}_T\cdot\boldsymbol{b_T}}}{(2\pi)^{2}} 
W^{S\!I\!D\!I\!S}(x,z,b_T,Q)+Y^{S\!I\!D\!I\!S}(x,z,q_T,Q)  
\,,\label{SIDIS-CSS}
\end{align}
where 
$q_T$ is the virtual photon momentum in the frame where the incident nucleon $N$ 
and the produced hadron $h$ are head to head, and 
\be
\sigma_0^{DIS}= \frac{4\pi\alpha^2_{\rm em}}{s x y^2} \left(1-y+\frac{y^2}{2}\right) \, , 
\ee
with the usual DIS kinematical variables $x=Q^2/(2P\cdot q)$, $y=P\cdot q/P\cdot l$.
Resummation is performed in the $b_T$ space, the Fourier conjugate of transverse momentum space, where momentum 
conservation laws can be taken into account more easily.
As mentioned above, the cross section is separated into two parts: a regular part, Y, and a resummed part, W.
Notice that, for SIDIS, we most commonly refer to the transverse momentum $\boldsymbol{P}_T$
of the final detected hadron, $h$, in the $\gamma^* N$ c.m. frame, rather than to the virtual photon 
momentum $\boldsymbol{q}_T$, in the $Nh$ c.m. frame. 
They are simply related by the hadronic   momentum fraction $z$ through the expression 
$\boldsymbol{P}_T=-z \,\boldsymbol{q}_T$, so that
\be
\frac{d\sigma}{d x\, d y\, d z\, d {P}_{T}^2}=\frac{d\sigma}{d x\,  d y \, d z\, d {q}_{T}^2}\frac{1}{z^2}\,.   
\ee

\subsection{The resummed term W \label{sec:Wterm}}

In the CSS resummation scheme, the term $W^{SIDIS}(x,z,b_T,Q)$, see Eq.~\eqref{SIDIS-CSS} resums the soft gluon contributions, large when $q_T \ll Q$: 
\begin{equation}
W^{S\!I\!D\!I\!S}(x,z,b_T,Q)= \exp\left[S_{pert}(b_T,Q)\right]\sum_j\! e_j^2\sum_{i,k}
\,C_{ji}^{\rm in}\otimes f_{i}(x,\mu_b^2)
\,C_{kj}^{\rm out}
\otimes D_{k}(z,\mu_b^2),
\label{eq:W}
\end{equation}	
where  $j=q,\bar q$ runs over all quark flavors available in the process, $i,k = q,\bar q, g$, and 
\begin{equation}
S_{pert}(b_T,Q)=-\int\limits_{\mu_b^2}^{Q^2}\frac{d \mup^2}{\mup^2}\left[A(\alpha_s(\mup))\ln\left(\frac{Q^2}{\mup^2}\right)+B(\alpha_s(\mup))\right] 
\label{S}
\end{equation}
is the perturbative Sudakov form factor. The intermediate scale $\mu_b(b_T)=C_1/b_T$ is chosen to optimize 
the convergence of the truncated perturbative series, $C_1=2\exp(-\gamma_E)$ and $\gamma_E$ is the Euler's constant. 
$A_j$ and $B_j$ are functions that can be expanded in series of $\alpha_s$, 
 \begin{eqnarray}
 A(\alpha_s(\mu))&=&\sum_{n=1}^{\infty}\left(\frac{\alpha_s}{\pi}\right)^n A^{(n)}\, ,\\
 B(\alpha_s(\mu))&=&\sum_{n=1}^{\infty}\left(\frac{\alpha_s}{\pi}\right)^n B^{(n)}\,,
  \end{eqnarray}
  and the coefficients $A^{(n)}$ and $B^{(n)}$ can be calculated in perturbative QCD. 
The symbol $\otimes$  in Eq.~(\ref{eq:W}) represents the usual collinear convolution of the Wilson 
coefficients $C_{ji}^{in}$, $C_{kj}^{out}$  and the collinear Parton Distributfion Functions (PDFs) $f_i(x,\mu_b^2)$, and collinear
 fragmentation functions (FF) $D_{k}(z,\mu_b^2)$.  
 \begin{eqnarray}
 C \otimes f(x) &\equiv&\int_{x}^{1} \frac{d \hat x}{\hat x}  C\left(\frac{x}{\hat{x}}\right) f(\hat x) \, .
 \label{eq:convolution}
 \end{eqnarray} 
Wilson coefficients $C$ are calculable in perturbative QCD; omitting parton indices one has
 \begin{eqnarray}
 C(x,\alpha_s(\mu_b))&=&\sum_{n=0}^{\infty}\left(\frac{\alpha_s(\mu_b)}{\pi}\right)^n C^{(n)}(x) \, .
 \label{eq:c}
 \end{eqnarray}
The theoretical error on the $q_T$ distributions depends on the accuracy to which perturbative coefficients are calculated:  
in particular, if one truncates the expansions at $A^{(1)}$ and $C^{(0)}$, then the resulting 
expression is at Leading Log (LL) accuracy, while  Next-to-Leading Log (NLL) accuracy is achieved 
by taking into account $A^{(1,2)}$, $B^{(1)}$ and $C^{(0,1)}$ coefficients ~\cite{Collins:1984kg,Davies:1984hs,Landry:2002ix,Koike:2006fn}:
\bea
\hspace*{-1.0cm}
A^{(1)} = C_F 
, \qquad
A^{(2)} =\frac{C_F}{2}\left[C_A\left(\frac{67}{18}-\frac{\pi^2}{6}\right)-\frac{10}{9}T_R \, n_f\right]
,\qquad
B^{(1)} = -\frac{3}{2} C_F 
,
\eea
where $C_F = 3/4$, $C_A =3$, $T_R = 1/2$, and $n_f$ is the number of active flavors.
Notice that, up to NLL, the coefficients A and B are process independent. For the Wilson coefficients we have \cite{Nadolsky:1999kb}: 
\begin{eqnarray}
C_{qq'}^{\rm(0)in}(x)&=&\delta_{qq'} \delta(1-x) \\
C_{qq'}^{\rm (0)out}(z)&=&\delta_{qq'} \delta(1-z) \\
C_{gq}^{\rm(0)out}(z) &=& C_{qg}^{\rm(0)in}(x)=0
\end{eqnarray}
\begin{eqnarray}
C_{qq'}^{\rm(1)in}(x)&=& \delta_{qq'}\frac{C_F}{2} \Big\{(1-x)-4 \delta(1-x) \Big\}\\
C_{qg}^{\rm(1)in}(x)&=&T_F[x(1-x)]\\
C_{qq'}^{\rm (1)out}(z)&=&\delta_{qq'}\frac{C_F}{2}\left\{(1-z)+2 \ln(z)\left[\frac{1+z^2}{1-z}\right] 
-4\delta(1-z)\right\} \\
C_{gq}^{\rm(1)out}(z)&=&\frac{C_F}{2}  \left\{z+2 \ln(z)\frac{1+(1-z)^2}{z}\right\}
\end{eqnarray}

The CSS formalism relies on a Fourier integral (\ref{SIDIS-CSS}) over $b_T$ which runs from zero to infinity. However, 
when $b_T$ is large one cannot rely completely on the perturbative computation of the corresponding coefficients.
The perturbative Sudakov factor, Eq.~(\ref{S}),  
hits the Landau pole in $\alpha_s$ at large values of 
$b_T$ (small values of $\mu_b$): this is a clear indication of non-perturbative physics. 
Predictions cannot be made without an ansatz prescription for the non-perturbative region, 
where $b_T$ is large. 
The CSS scheme, therefore, introduces a prescription which prevents $b_T$ from getting any larger than some 
(predefined) maximum value 
$b_{max}$: 
\begin{equation}
 b_*=\frac{b_T}{\sqrt{1+b_T^2/b_{max}^2}}\,.
 \label{b*}
\end{equation}
Accordingly, in the definition of $S_{pert}$,  $\mu_b(b_T)$ is replaced by $\mu_b(b_*)=C_1/b_*$. 

Notice that, for large values of $b_{max}$, $\mu_b = C_1/b_*$ tends to become smaller than the 
minimum scale available for the corresponding collinear parton distribution/fragmentation functions: 
in order to reliably use the collinear PDFs, in this case we freeze its value at $1.3$ GeV.

Then the cross section is written as
\bea
\frac{d\sigma^{total}}{d x \, d y\, d z\,   d{q}_{T}^2} &=& \pi \sigma_0^{D\!I\!S} \!\!\int\limits_{0}^{\infty}
\frac{d {b}_T b_T}{(2\pi)} J_0( {q}_T {b_T})
W^{S\!I\!D\!I\!S}(x,z,b_*,Q)\exp\left[S_{N\!P}(x,z,b_T,Q)\right] \nonumber \\
&+&Y(x,z,q_T,Q) \, , \label{SIDIS-CSS1}
\eea
where $W^{SIDIS}$ is now evaluated at  $b_T=b_*$, while  $S_{N\!P}(x,z,b_T,Q)$ is a new function which accounts for 
the non-perturbative behaviour of the cross section at large $b_T$. Clearly, $S_{N\!P}$ should be equal to zero when $b_T = 0$. 

The predictive power of the $b_T$-space resummation formalism is limited by our inability to calculate the non-perturbative 
distributions at large $b_T$. However, most of these non-perturbative distributions are believed to be universal  
and can be extracted from experimental data on different processes and allow for predictions for other 
measurements. Non-perturbative physics is also interesting as it gives us insights on fundamental properties 
of the nucleon.

As already mentioned, the results of our studies can be easily extended to the Collins TMD evolution scheme~\cite{Collins:2011zzd}.
In  Appendix~\ref{sec:TMD} we show that the two formalisms are equivalent to first loop.

\subsection{The Y-term\label{sec:Yterm}}

The resummed  
cross section, $W$, cannot describe the whole  $q_T$  range: 
it sums the logarithmic terms dominating the low $q_T$ region, 
but it does not include contributions to the total cross section which are less singular than $1/q_T^2$, that become important at large $q_T$. 
  Leaving out these terms introduces a relative error of ${\cal O} (q_T^2/Q^2)$, thus the resummed result is valid only if $q_T \ll Q$.
Ultimately, these terms are contained inside the Y-factor, which we are now going to define. 

The Next to Leading Order (NLO)\footnote{Notice that here NLO means first order in $\alpha_s$ of the collinear perturbative QCD cross 
section.}  cross section can be separated into an ``asymptotic part'', 
$d\sigma^{ASY}$, which includes all the contributions proportional to 
${Q^2}/{q_T^2}$ and to ${Q^2}/{q_T^2} \ln({Q^2}/{q_T^2})$, badly divergent at small $q_T$, and a regular part 
$Y^{SIDIS}(x,z,q_T,Q)$, 
the Y-term 
which includes all terms of the cross section which are, at most, logarithmic as $q_T\to 0$
and ensures a smooth transition of the cross section to the region of large $q_T$, so that
\be
\frac{d\sigma^{NLO}}{d x\,  d y\, d z \, d{q}_{T}^2} =\frac{d\sigma^{ASY}}{d x \, d y \, d z\, d 
{q}_{T}^2} + Y\,,
\ee
and inverting
\be
Y = \frac{d\sigma^{NLO}}{d x\,  d y\, d z \,d {q}_{T}^2} - \frac{d\sigma^{ASY}}{d x \, d y \,d z\, d{q}_{T}^2} \,.
\label{eq:yfirst}
\ee
The explicit expressions of $d\sigma^{NLO}$ and $d\sigma^{ASY}$ are given in Ref.~\cite{Koike:2006fn}.
In the CSS scheme~\cite{Collins:1984kg}, the diverging terms in the asymptotic part are then resummed so that the final cross section is given by 
 Eq.~(\ref{SIDIS-CSS}). 
 
\begin{figure}[t]
\centerline{
\includegraphics[width=7.4cm]{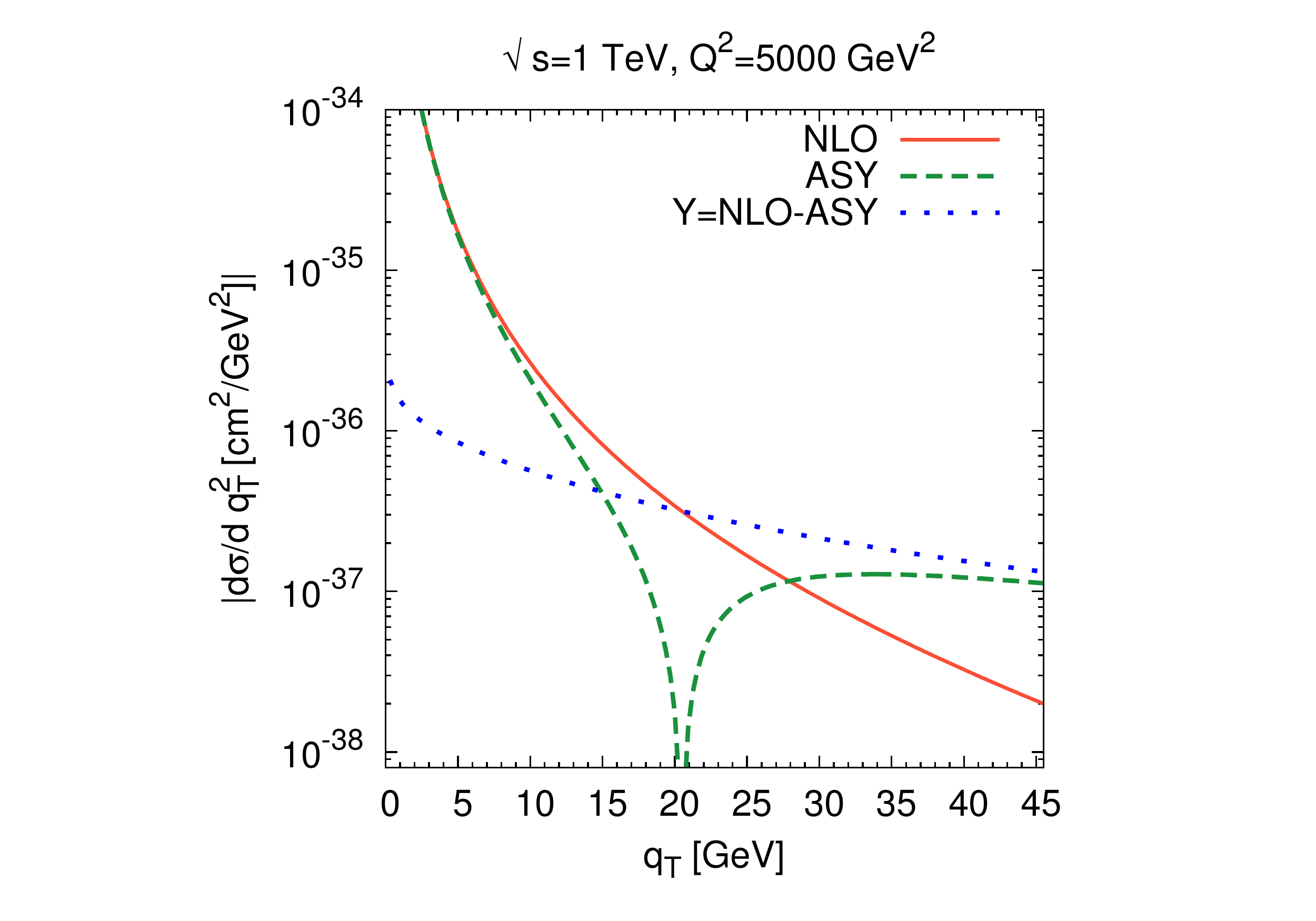}\hspace*{-2.5cm}
\includegraphics[width=7.4cm]{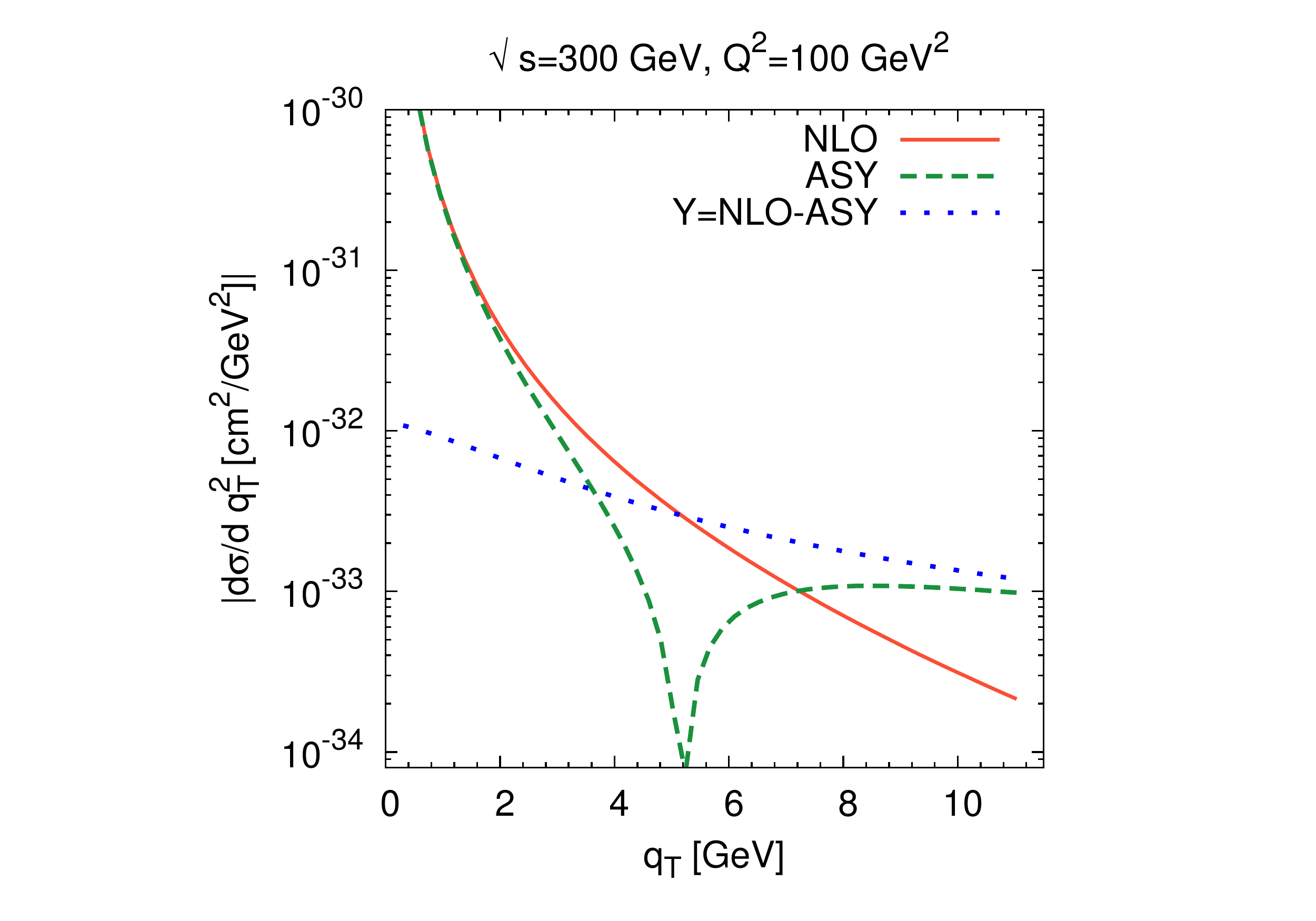} \hspace*{-2.5cm}
\includegraphics[width=7.4cm]{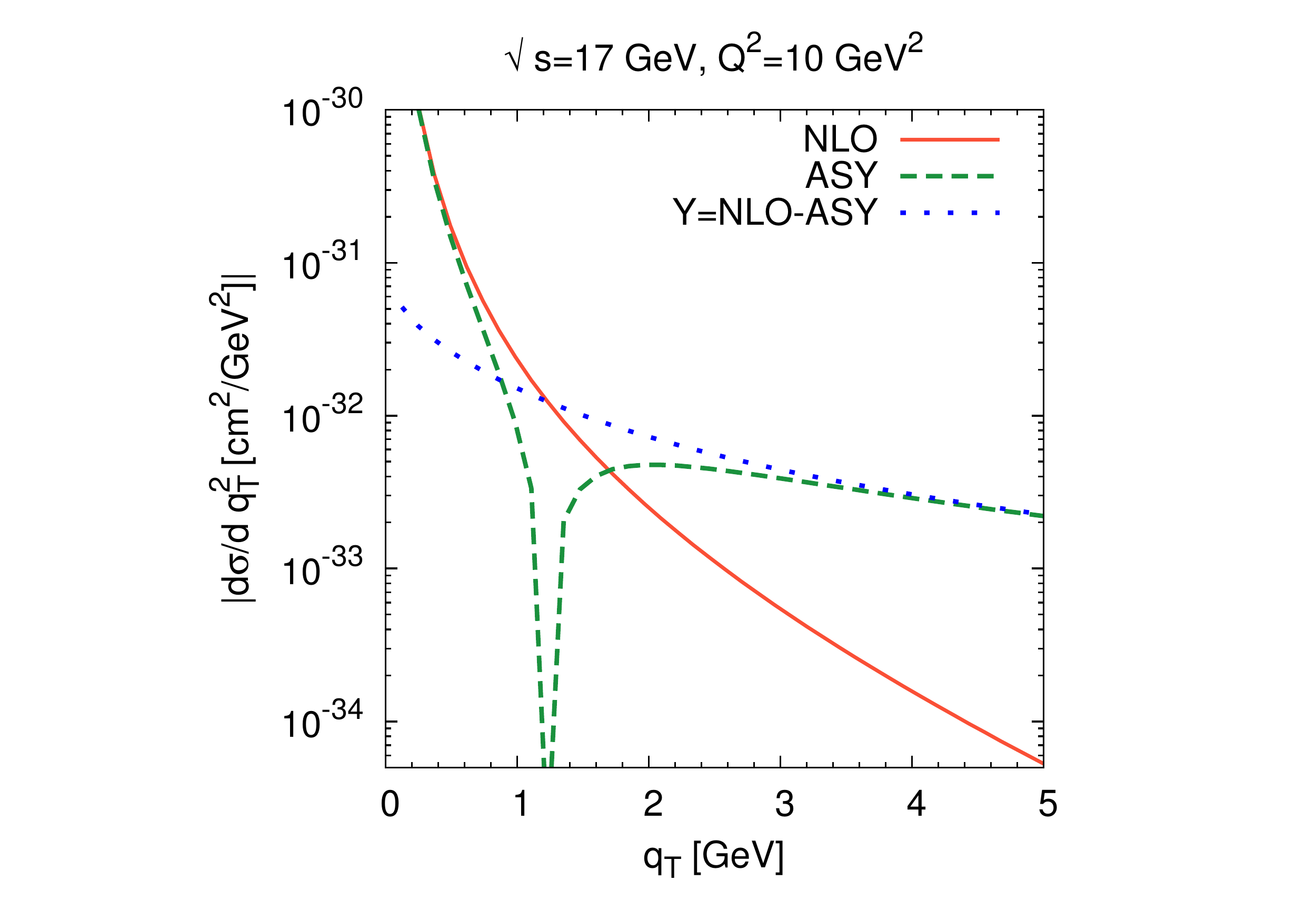}}
\vspace*{-8pt}
\caption{ Perturbative contributions to the SIDIS cross sections, $d\sigma^{ASY}$, $d\sigma^{NLO}$ and $Y$ factor, 
corresponding to three different SIDIS kinematical configurations: 
on the left panel $\sqrt{s}=1$ TeV, $Q^2=5000$ GeV$^2$, $x=0.055$ and $z=0.325$; 
on the central panel a HERA-like experiment with  $\sqrt{s}=300$ GeV, $Q^2=100$ GeV$^2$, $x=0.0049$ and $z=0.325$; 
on the right panel, a COMPASS-like experiment with $\sqrt{s}=17$ GeV, $Q^2=10$ GeV$^2$, $x=0.055$ and $z=0.325$.  
\label{f1}}
\end{figure}
%
Fig.~\ref{f1} shows the $d\sigma^{ASY}$, $d\sigma^{NLO}$ and Y cross section contributions for  SIDIS $\pi^+$ production off a proton target:
the left panel corresponds to an extremely high energy SIDIS experiment with $\sqrt{s}=1$~TeV, $Q^2=5000$~GeV$^2$, $x=0.055$ and $z=0.325$;
in the central panel we choose an intermediate,  
HERA-like kinematics configuration, with $\sqrt{s}=300$ GeV, $Q^2=100$ GeV$^2$, $x=0.0049$ and $z=0.325$; the right panel corresponds 
to a lower energy SIDIS experiment like COMPASS, 
with $\sqrt{s}=17$~GeV, $Q^2=10$~GeV$^2$, $x=0.055$ and $z=0.325$. In our study we use the MSTW08 PDF set~\cite{Martin:2009iq} and the DSS FF set~\cite{deFlorian:2007aj}.

Notice that at large $q_T$ 
$d\sigma^{ASY}$ becomes negative and therefore unphysical 
(we show the absolute value of the asymptotic NLO cross section in Fig.~\ref{f1} as a dashed, green line). 
Consequently, the $Y=d\sigma^{NLO}-d\sigma^{ASY}$ term can become much larger than the $NLO$ cross section in that region.

\section{Matching prescriptions\label{sec:match}}

One of the underlying ideas of the standard resummation scheme is that 
the resummed cross section has to be matched, at some point, to the fixed order 
cross section.

\noindent
By defining 
\be
W = \pi \sigma_0^{D\!I\!S} \!\!\int\limits_{0}^{\infty}\frac{d {b}_T b_T}{(2\pi)} J_0( {q}_T {b_T})
W^{S\!I\!D\!I\!S}(x,z,b_T,Q)\,,
\ee
and neglecting (for the moment) non-perturbative contributions, the final cross section can be written in a short-hand notation as 
 \be
d\sigma^{total} = W + Y  \,.
\label{eq:yterm2}
\ee
In the region where $q_T \simeq Q$,  
the logarithmic terms are expected to be small so, in principle, the resummed cross section should be equal or very similar to its 
asymptotic counterpart, $d\sigma^{ASY}$. Therefore, the cross section in Eq.~(\ref{eq:yterm2}) should almost exactly  
{\it match} the NLO cross section, $d\sigma^{NLO}$:
\be
d\sigma^{total} = W + Y \xrightarrow{q_T\sim \,Q}  d\sigma^{ASY} + Y = d\sigma^{ASY} + d\sigma^{NLO} - d\sigma^{ASY} = d\sigma^{NLO}\,.
\label{eq:yterm}
\ee
It is crucial to stress that this matching prescription at $q_T \simeq Q$ only works if $W \simeq d\sigma^{ASY}$
over a non-negligible range of $q_T$ values, as the matching should be {\it smooth} as well as continuous. 

At small $q_T$, one expects that $d\sigma^{ASY}$ and $d\sigma^{NLO}$ 
are dominated by the same diverging terms, proportional to ${Q^2}/{q_T^2}$ and to ${Q^2}/{q_T^2} \ln({Q^2}/{q_T^2})$; 
therefore, they should almost cancel in the definition of $Y$ leaving in $d\sigma^{total}$ the sole resummed cross section W
\be
d\sigma^{total} = W + Y \xrightarrow{q_T\ll \,Q}  W\,.
\label{eq:yterm1}
\ee
This cancellation occurs only as long as we keep away from the singularity in $Y$, at  $q_T=0$.
Thus, 
this matching prescription is such that the total cross section is dominated by W at small $q_T$, and by  $d\sigma^{NLO}$ at large $q_T$. 
In the intermediate $q_T$ region, it is given by the sum $(W+Y)$, Eq.~\eqref{eq:yterm2}.

%
\begin{figure}[t]
\centerline{
\includegraphics[width=7.4cm]{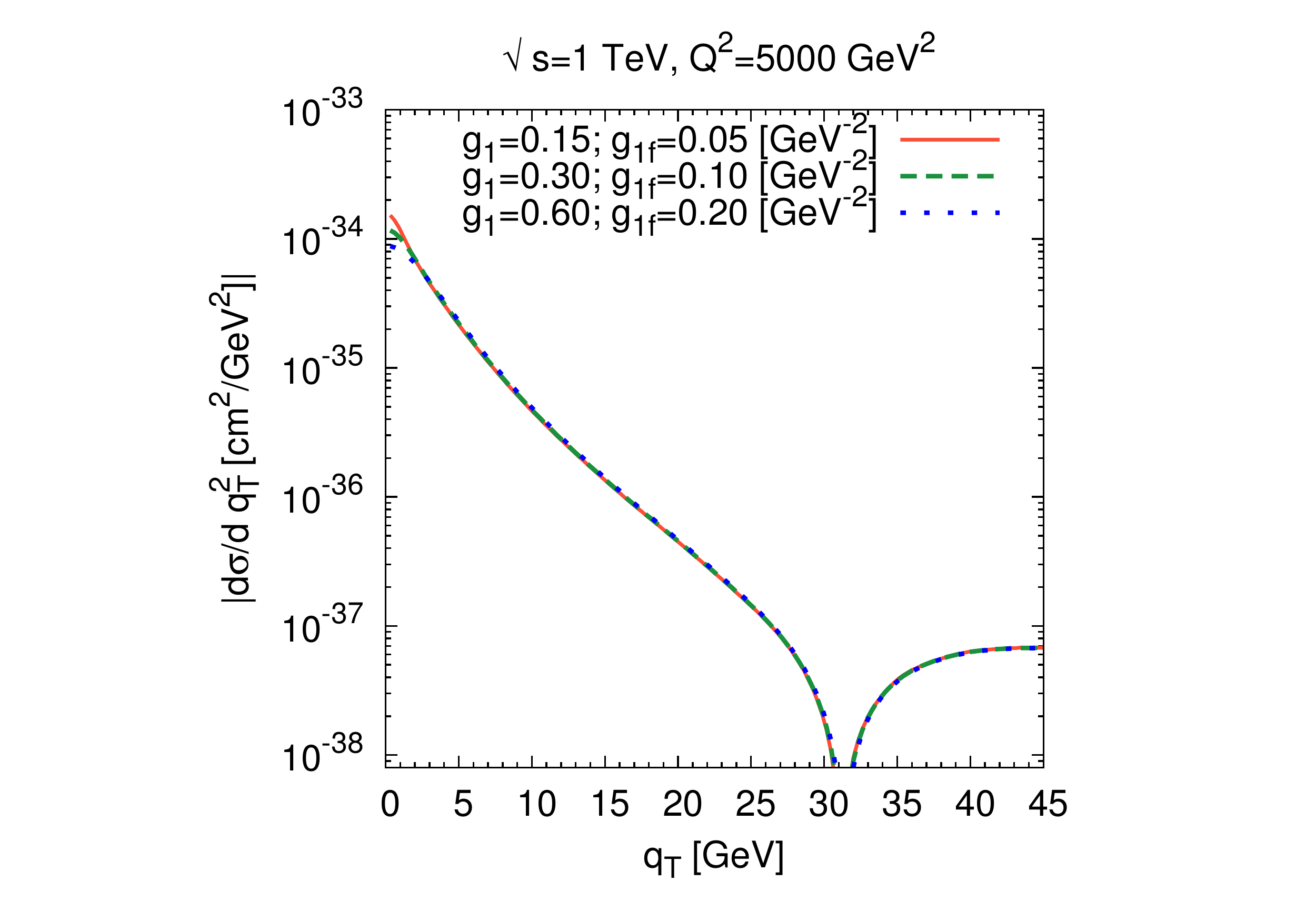} \hspace*{-2.5cm}
\includegraphics[width=7.4cm]{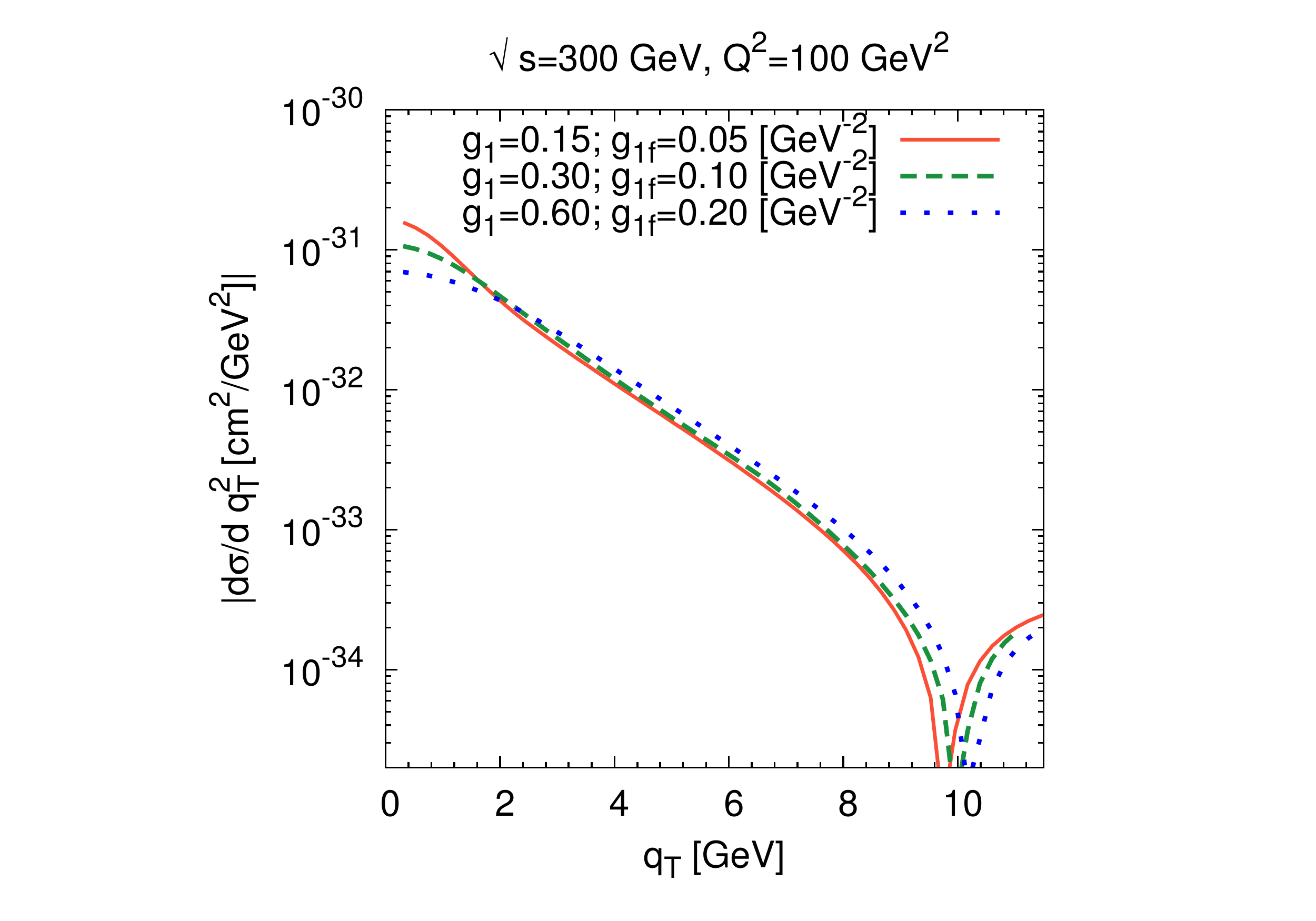} \hspace*{-2.5cm}
\includegraphics[width=7.4cm]{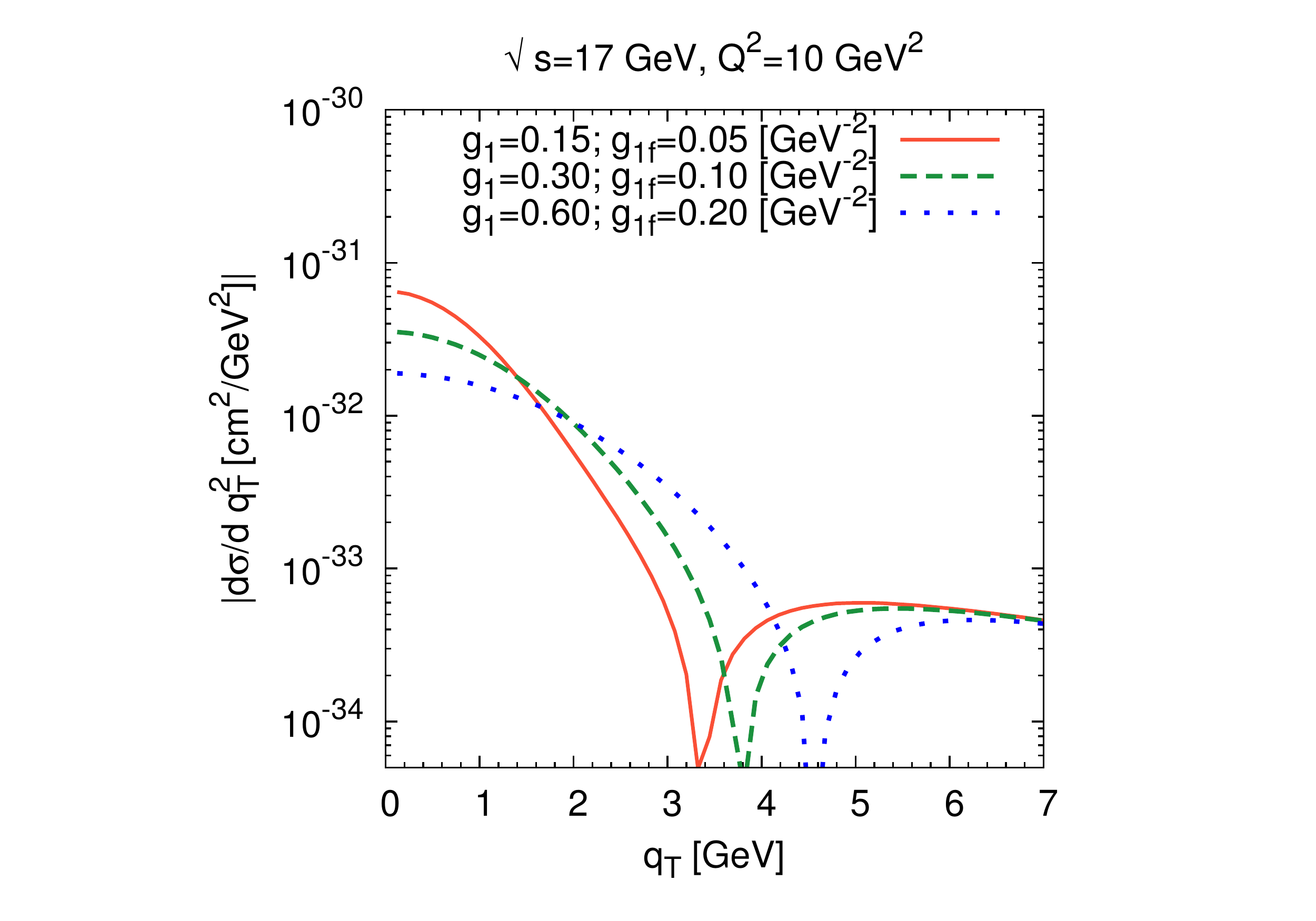}} 
\vspace*{-8pt}
\caption{Resummed term of the SIDIS cross section including the non-perturbative contribution $S_{N\!P}$ in the Sudakov factor, 
calculated at three different values of $g_1$ and $g_{1f}$ and corresponding to the three 
different SIDIS kinematical configurations defined in Fig.~\ref{f1}.
Here $b_{max}= 1.0$ GeV$^{-1}$. \label{f2}}
\end{figure}
%

\subsection{Non-perturbative contribution to the Sudakov factor \label{sec:NP}}

At this stage, one should wonder whether, given a well-defined SIDIS scattering process, a kinematical range in 
which $W \simeq d\sigma^{ASY}$ actually does exist, where the matching can successfully be performed.
To answer this question we need to compute the W-term, which necessarily implies specifying its non-perturbative 
behaviour. 
The considerations of 
Eq.~\eqref{eq:yterm} are based on the assumption that non-perturbative contributions do {\it not} affect the numerical 
calculations. To check this assumption,  
let us choose a particular value $b_{max}= 1.0$ GeV$^{-1}$ and consider a simple model for the non-perturbative function $S_{N\!P}$:
\be
S_{N\!P} =  \left(- \frac{g_1}{2}  - \frac{g_{1f}} {2 z^2} - g_2 \ln\left(\frac{Q}{Q_0}\right)\right) b_T^2  \,.
\ee
The actual values of these parameters are not important for our studies and 
the conclusions may well hold for different choices of the parameters. 
Here we set $g_2 = 0$ (GeV$^2$) in order not to enter into the details of  the exact functional form 
of $S_{NP}$, which have no influence.

We now define as $W^{NLL}$ the NLL resummed cross section which includes the non-perturbative Sudakov factor
\be
W^{NLL} = \pi \sigma_0^{D\!I\!S} \!\!\int\limits_{0}^{\infty}\frac{d {b}_T b_T}{(2\pi)} J_0( {q}_T {b_T})
W^{S\!I\!D\!I\!S}(x,z,b_*,Q)\exp\left[S_{N\!P}(x,z,b_T,Q)\right]\,,
\label{WNLL}
\ee 
with $W^{S\!I\!D\!I\!S}(x,z,b_*,Q)$ of Eq.~\eqref{eq:W} calculated at NLL order as explained in Section~\ref{resummation}.

Obviously, having introduced a parametrization to represent $S_{N\!P}$, our results will now inevitably be affected 
by some degree of model dependence, according to the kinematics of the SIDIS 
process under consideration. 
Fig.~\ref{f2} shows the resummed term of the SIDIS cross section, including the non-per\-tur\-ba\-tive 
contribution to the Sudakov factor, 
$S_{N\!P}$, 
calculated with three different values of the pair ($g_1 , g_{1f}$), and corresponding to the same three
different SIDIS kinematical configurations considered in Fig.~\ref{f1}.
These plots clearly show that, while in an extremely high energy and $Q^2$ configuration (left panel) 
the dependence on the non-perturbative parameters is limited to the region of very small $q_T$, at intermediate 
energies (central panel) the non-per\-tur\-ba\-tive content of the Sudakov factor, $S_{N\!P}$, 
induces a sizable dependence on the parameters of the model over the whole $q_T$ range. 
At smaller energies and $Q^2$ (right panel), 
the dependence of the SIDIS cross section 
on the value of the non-perturbative parameters is extremely strong, and the three curves change sign at very 
different values of $q_T$.  Therefore, in this case, we cannot expect a successful cancellation between $d\sigma^{ASY}$ and $W^{NLL}$.

\begin{figure}[t]
\centerline{
\includegraphics[width=7.4cm]{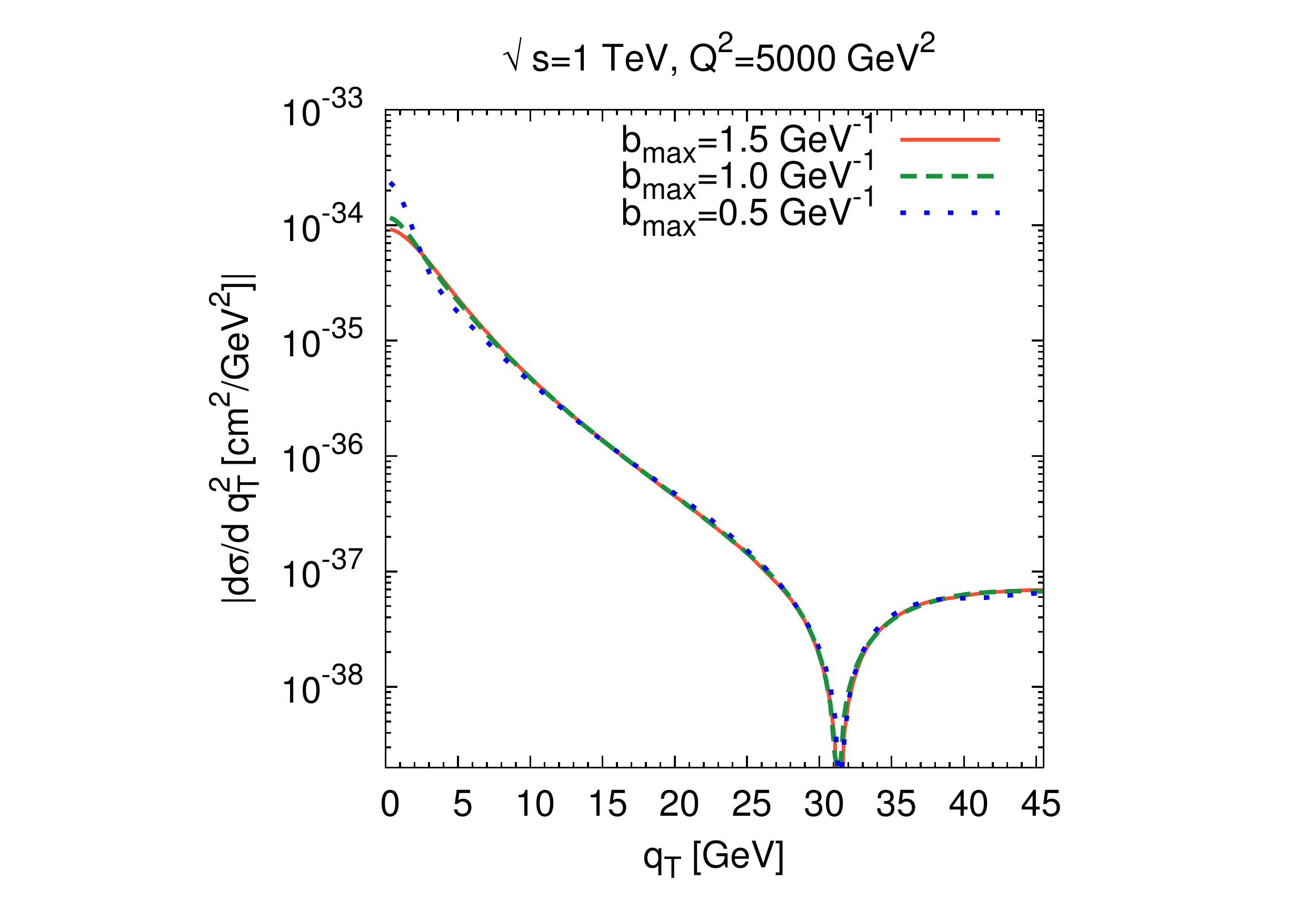}\hspace*{-2.5cm}
\includegraphics[width=7.4cm]{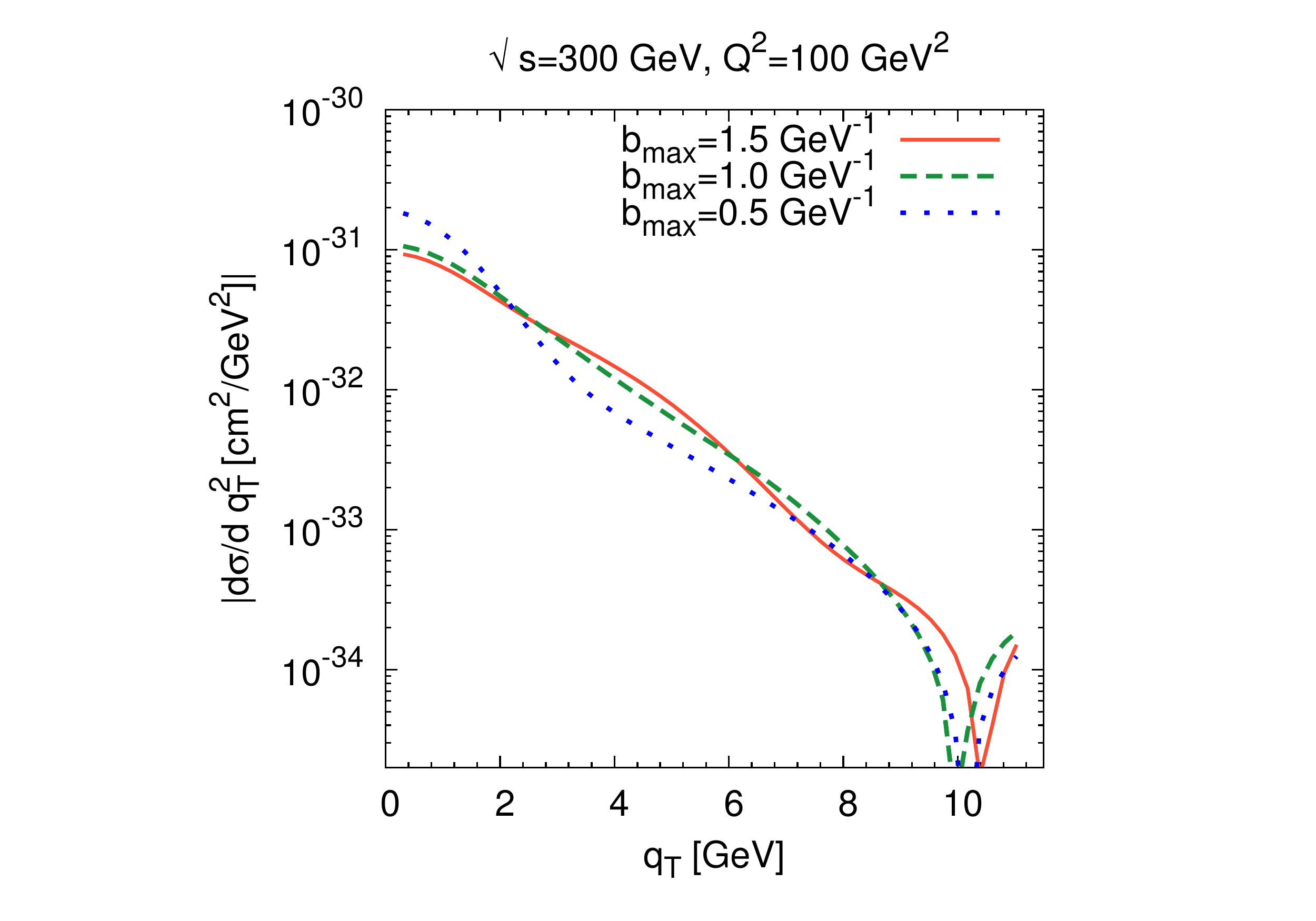}\hspace*{-2.5cm}
\includegraphics[width=7.4cm]{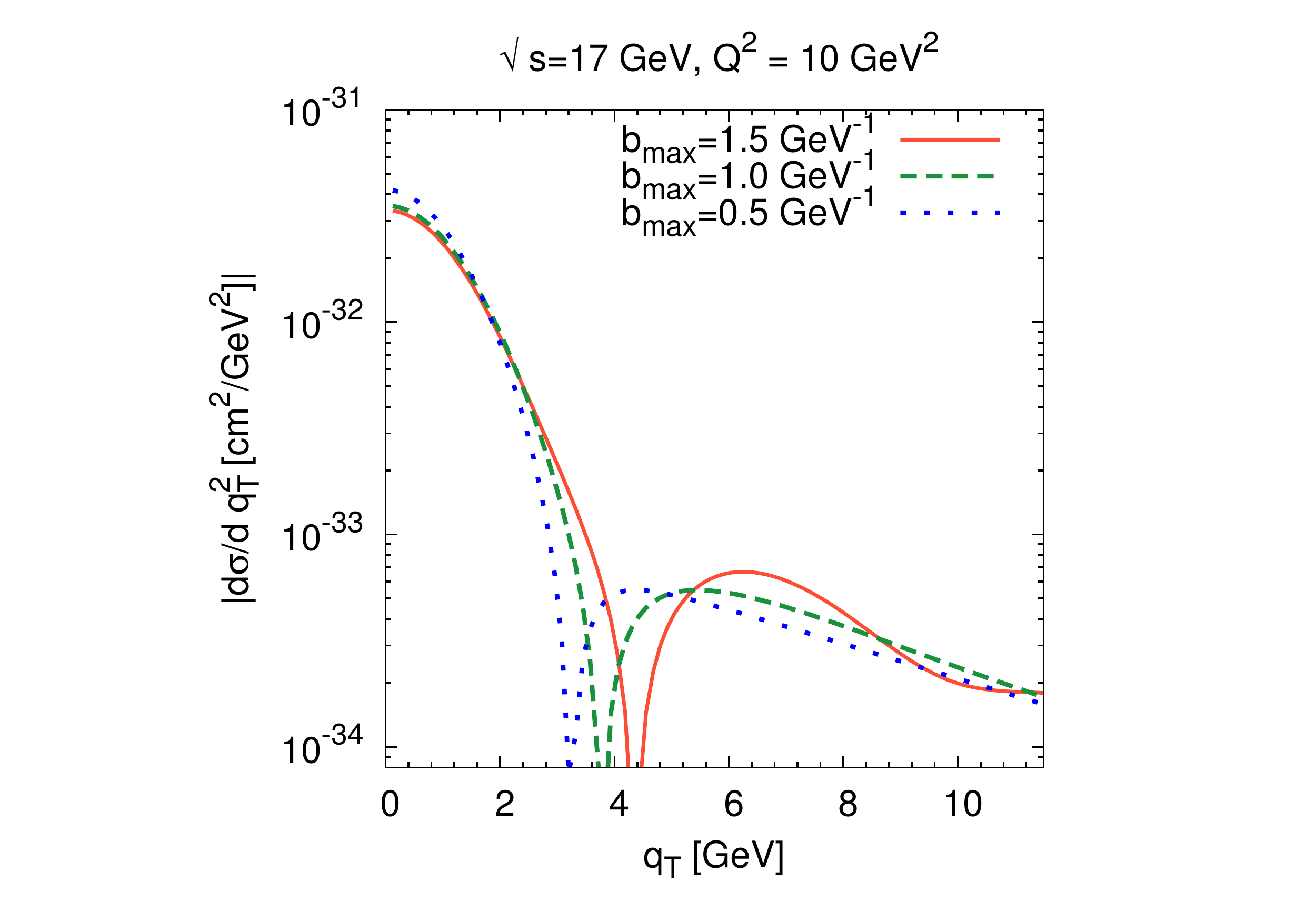} }
\vspace*{-8pt}
\caption{  
The resummed cross section $W^{NLL}(q_T)$ corresponding to the three
different SIDIS kinematical configurations defined in Fig.~\ref{f1}.
Here $b_{max}$ varies from $1.5$ GeV$^{-1}$ to $0.5$ GeV$^{-1}$, while $g_1$ and $g_{1f}$ are fixed at 
$g_1=0.3$ GeV$^2$, $g_{1f}=0.1$~GeV$^2$.
\label{f3}}
\end{figure}

\begin{figure}[t]
\centerline{
\includegraphics[width=6.7cm]{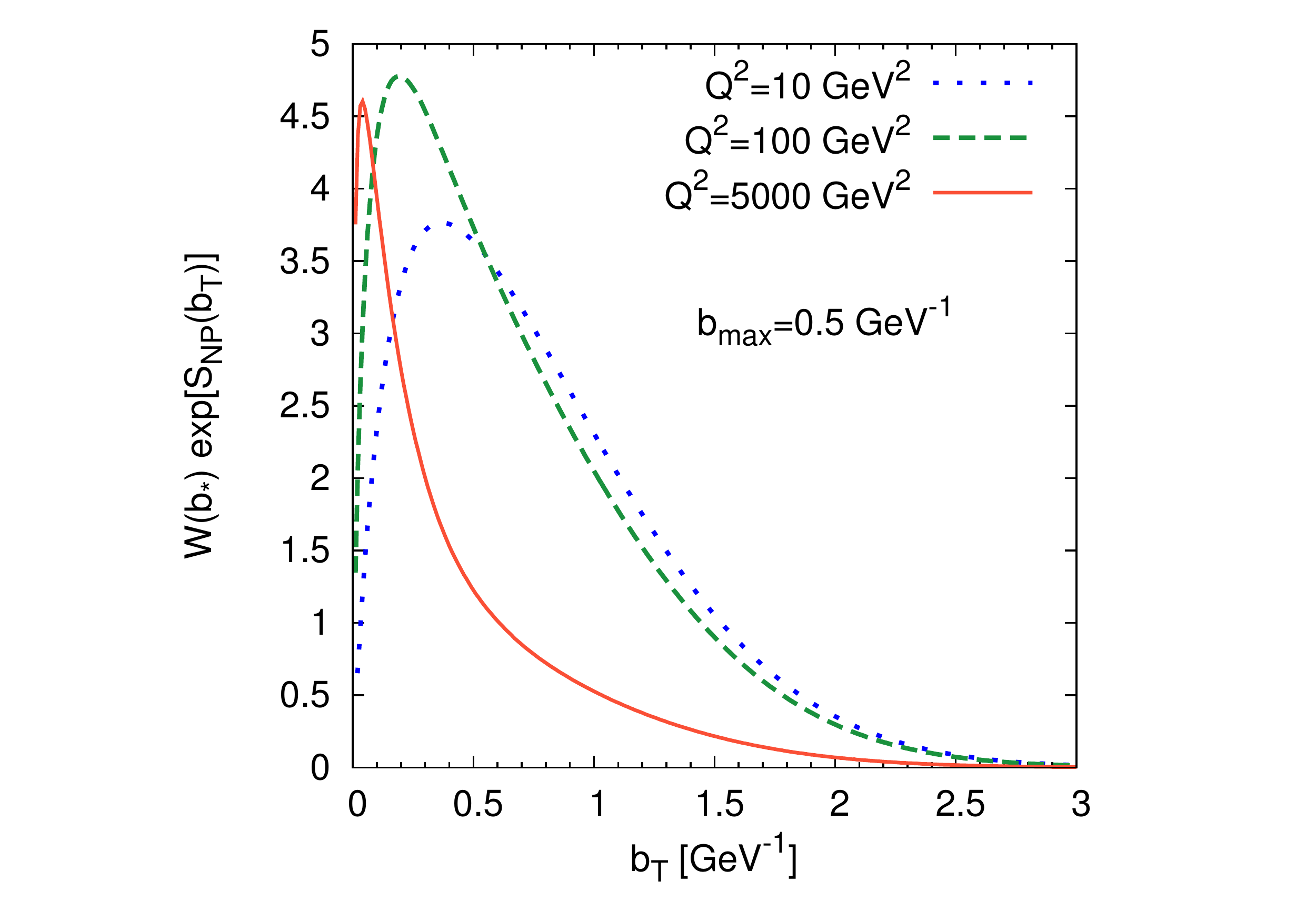}\hspace*{-1.8cm}
\includegraphics[width=6.7cm]{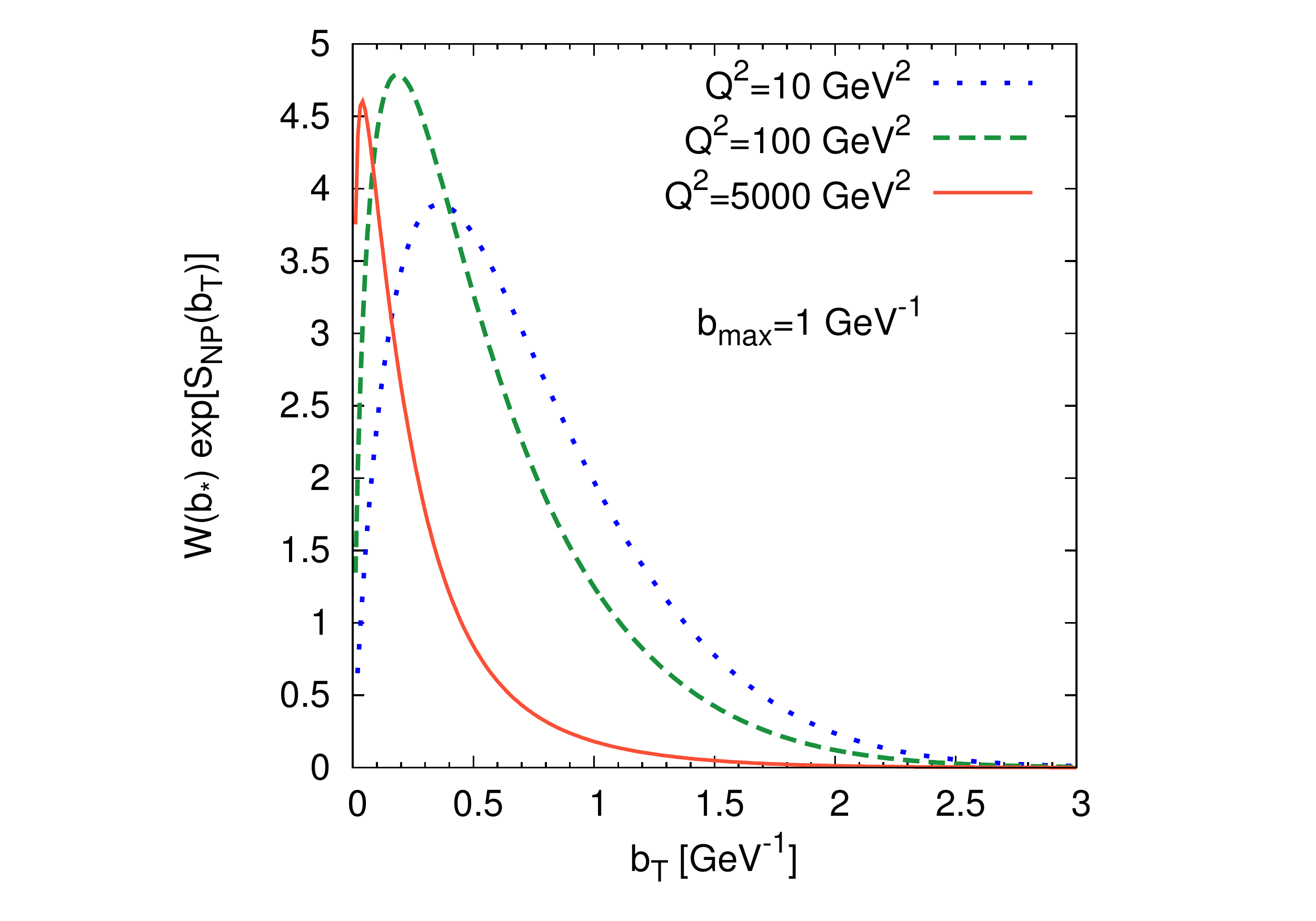} \hspace*{-1.8cm}
\includegraphics[width=6.7cm]{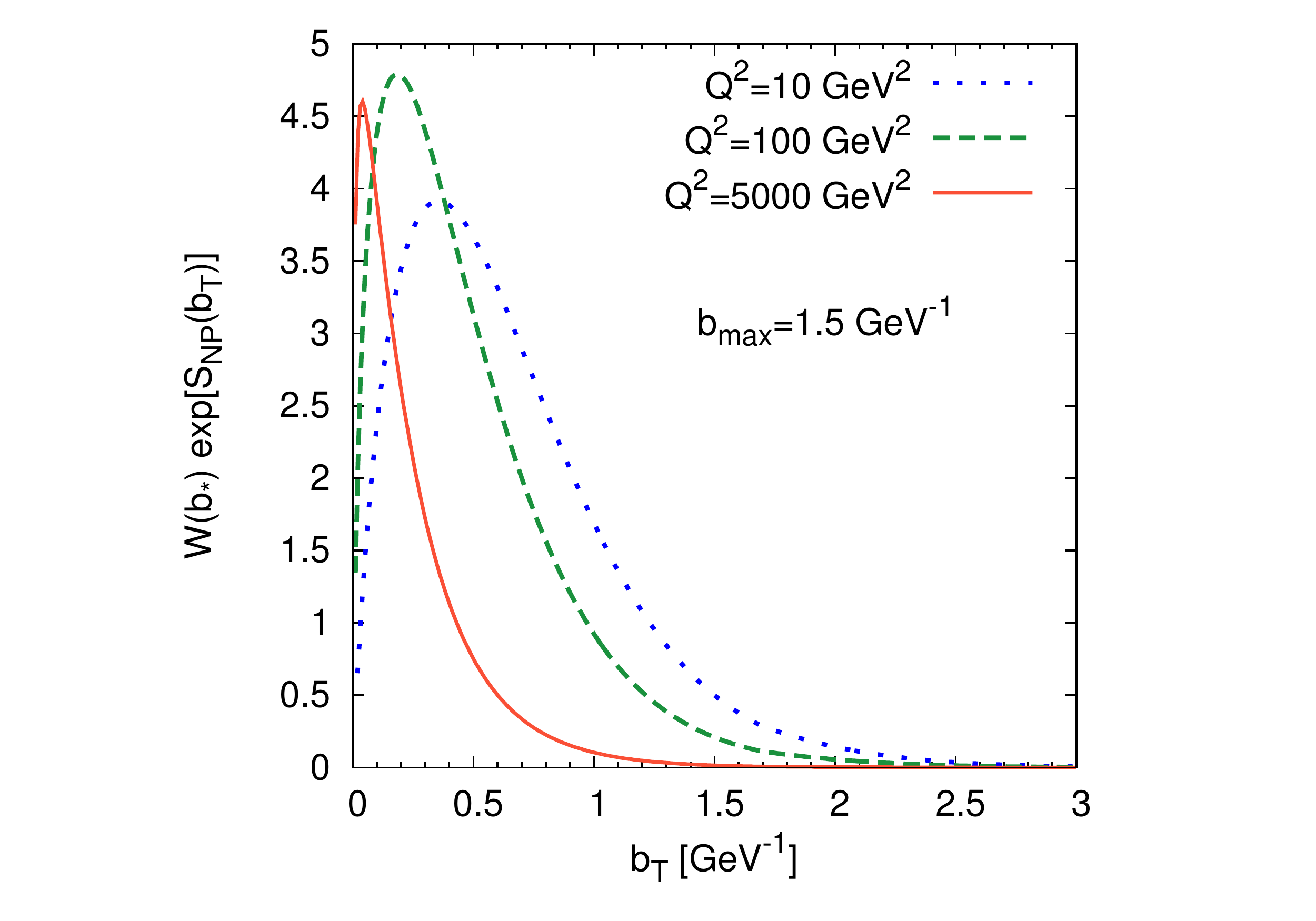} }
\vspace*{-8pt}
\caption{ 
The resummed term $W^{SIDIS}(b_*)\exp[S_{NP}(b_T)]$ as a function of $b_T$ corresponding to three 
different SIDIS kinematical configurations, $Q^2=5000$ GeV$^2$,  $Q^2=100$ GeV$^2$,  
and $Q^2=10$ GeV$^2$. Here $b_{max}$ varies from $0.5$ GeV$^{-1}$ (left panel) to  $1$ GeV$^{-1}$ (central panel), 
$1.5$ GeV$^{-1}$ (right panel). In order to compare different kinematical configurations, in this plot we fix 
$x$ and $z$ to values compatible with all of them: $x=0.055$ and $z=0.325$.
\label{f4}}
\end{figure}

\subsection{Dependence of the total cross section on the $b_{max}$ parameter\label{sec:bmax}}

As mentioned in Section~\ref{resummation}, the parameter $b_{max}$ controls the $b_T$ scale of transition between perturbative 
and non-perturbative regimes, see Eqs.~\eqref{b*} and~\eqref{SIDIS-CSS1}, by limiting the value of $b_T$ to the point in which 
perturbative calculations reach the boundary of their validity.
It is therefore very interesting to study the influence of the choice of $b_{max}$ on the cross section, at fixed values of 
the non-perturbative parameters $g_1$ and $g_{1f}$. 
In Fig.~\ref{f3} we plot the resummed cross section of Eq.~(\ref{SIDIS-CSS1}) at three different values of 
$b_{max}=$   $1.5$ GeV$^{-1}$, $1.0$ GeV$^{-1}$ and  $0.5$ GeV$^{-1}$, having fixed $g_1=0.3$ GeV$^2$, $g_{1f}=0.1$ GeV$^2$.
By comparing the plots, from right to left, we notice that in the COMPASS case there is a strong dependence on the chosen value of 
$b_{max}$ and the non-perturbative contribution dominates almost 
over the entire range. In the HERA-like kinematics we observe a slightly milder, but still sizable, residual dependence 
on $b_{max}$, even at large $q_T$. Ultimately, it is only when we reach the highest energies and $Q^2$ values of the leftmost 
plot that we find an almost complete insensitiveness to the chosen value of $b_{max}$.

To understand this effect, we can study the behaviour of 
$W^{S\!I\!D\!I\!S}(b_*)\exp\left[S_{N\!P}(b_T)\right]$, as a function of $b_T$. Fig.~\ref{f4} shows that  
these $b_T$ distributions, as expected, become increasingly peaked and narrow as $Q^2$ grows, 
reflecting the dominance of smaller and smaller $b_T$ contributions at growing energies and $Q^2$: clearly, for the COMPASS 
kinematics (dotted-blue line), the integrand shows a wider $b_T$ distribution, 
with a larger tail, compared to that corresponding to higher energies and larger $Q^2$ configurations
(dashed-green line and 
solid-red line).

\begin{figure}[b]
\centerline{
\includegraphics[width=7.4cm]{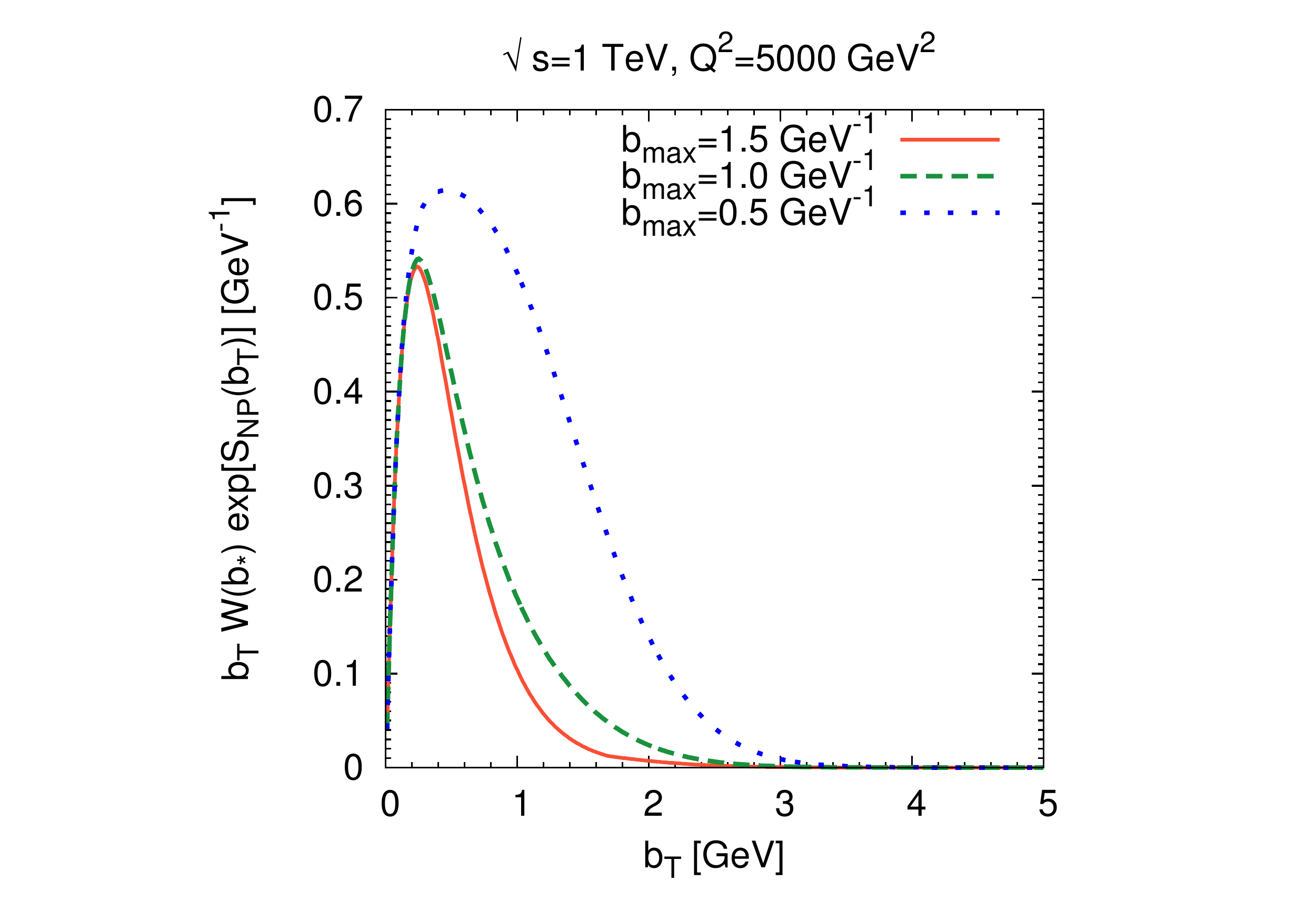} \hspace*{-2.5cm}
\includegraphics[width=7.4cm]{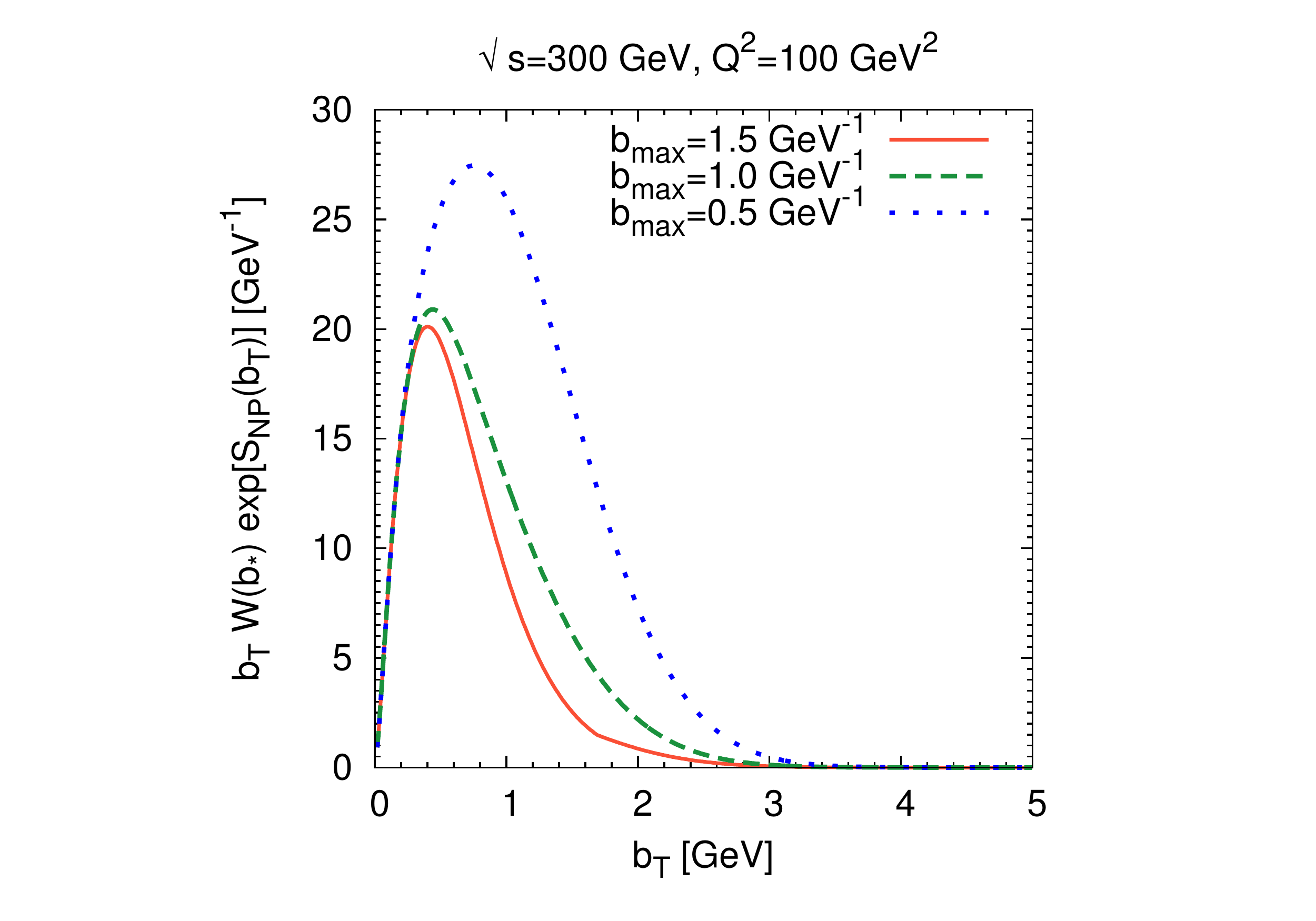} \hspace*{-2.5cm}
\includegraphics[width=7.4cm]{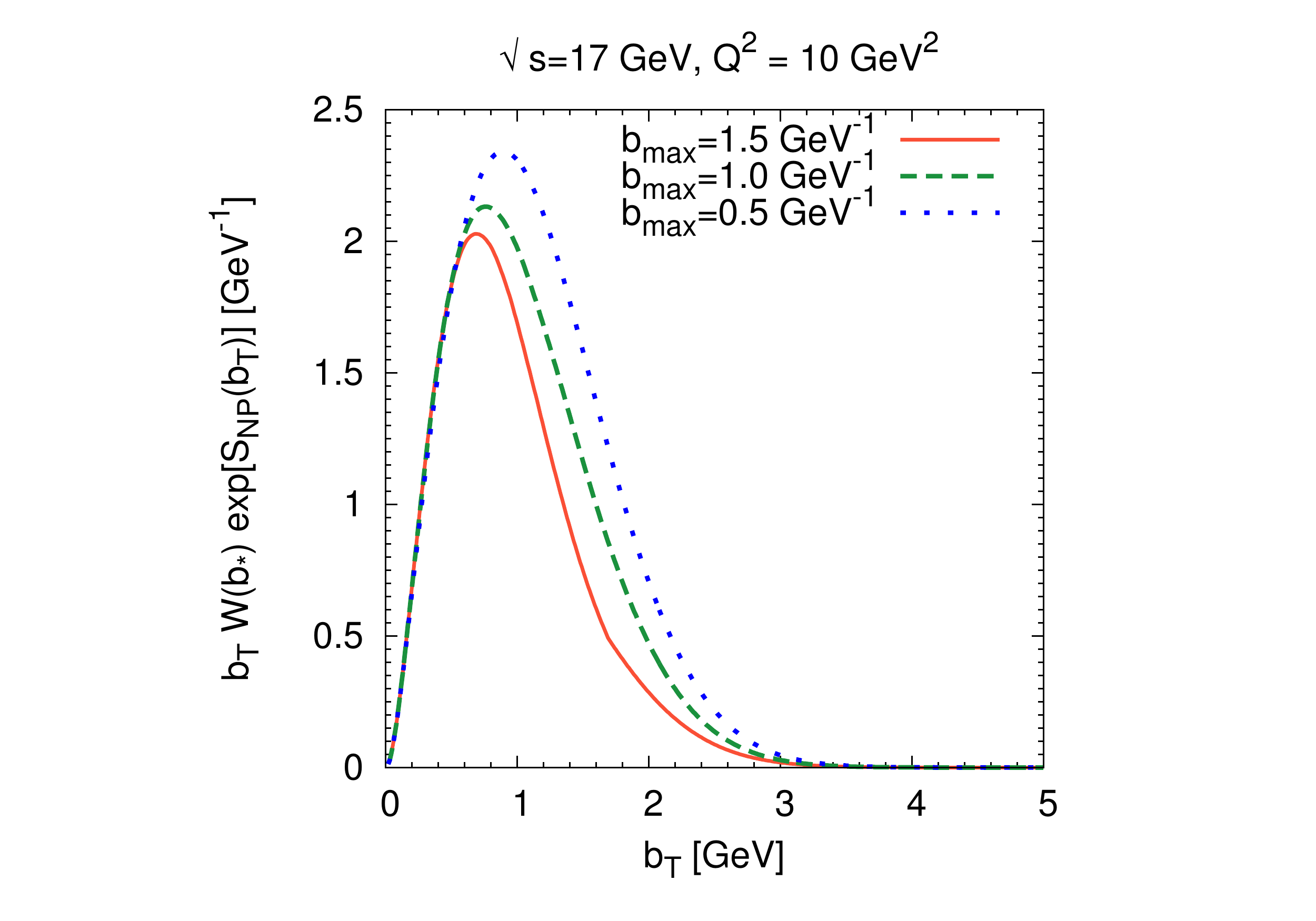} } 
\vspace*{-8pt}
\caption{ 
The resummed term $ b_T W^{SIDIS}(b_*)\exp[S_{NP}(b_T)]$ corresponding to the three
different SIDIS kinematical configurations defined in Fig.~\ref{f1}.
Here $b_{max}$ varies from $1.5$ GeV$^{-1}$(solid line) to  $1$ GeV$^{-1}$(dashed line), $0.5$ GeV$^{-1}$ (dotted line).
\label{f5}}
\end{figure}

From Fig.~\ref{f5}, where we plot  the integrand of Eq.~\eqref{WNLL}, $b_T \,W^{SIDIS}(b_*)\exp[S_{NP}(b_T)]$, 
one can learn about the dependence on the 
choice of $b_{max}$: 
at each fixed kinematical configuration, the peak moves toward larger values as $b_{max}$ decreases. 
Moreover, Fig.~\ref{f5} shows how the tail behaviour is affected by different choices of $b_{max}$:
in fact, as  $b_{max}$ fixes the $b_T$ scale of the transition between perturbative and non-perturbative regimes, 
the distributions obtained from growing 
values of $b_{max}$ die faster in $b_T$, because the non-perturbative 
contribution sets in at larger and larger values of $b_T$.

\subsection{Y term matching \label{sec:matchY}}

It should now be clear that a successful matching heavily depends on the subtle interplay between 
perturbative and non-perturbative contributions to the total cross section, and that finding a 
kinematical range in which the resummed cross section $W$ matches its asymptotic 
counterpart $d\sigma^{ASY}$, in the region $q_T\sim Q$, cannot be taken for granted. 

\begin{figure}[t]
\centerline{
\includegraphics[width=7.4cm]{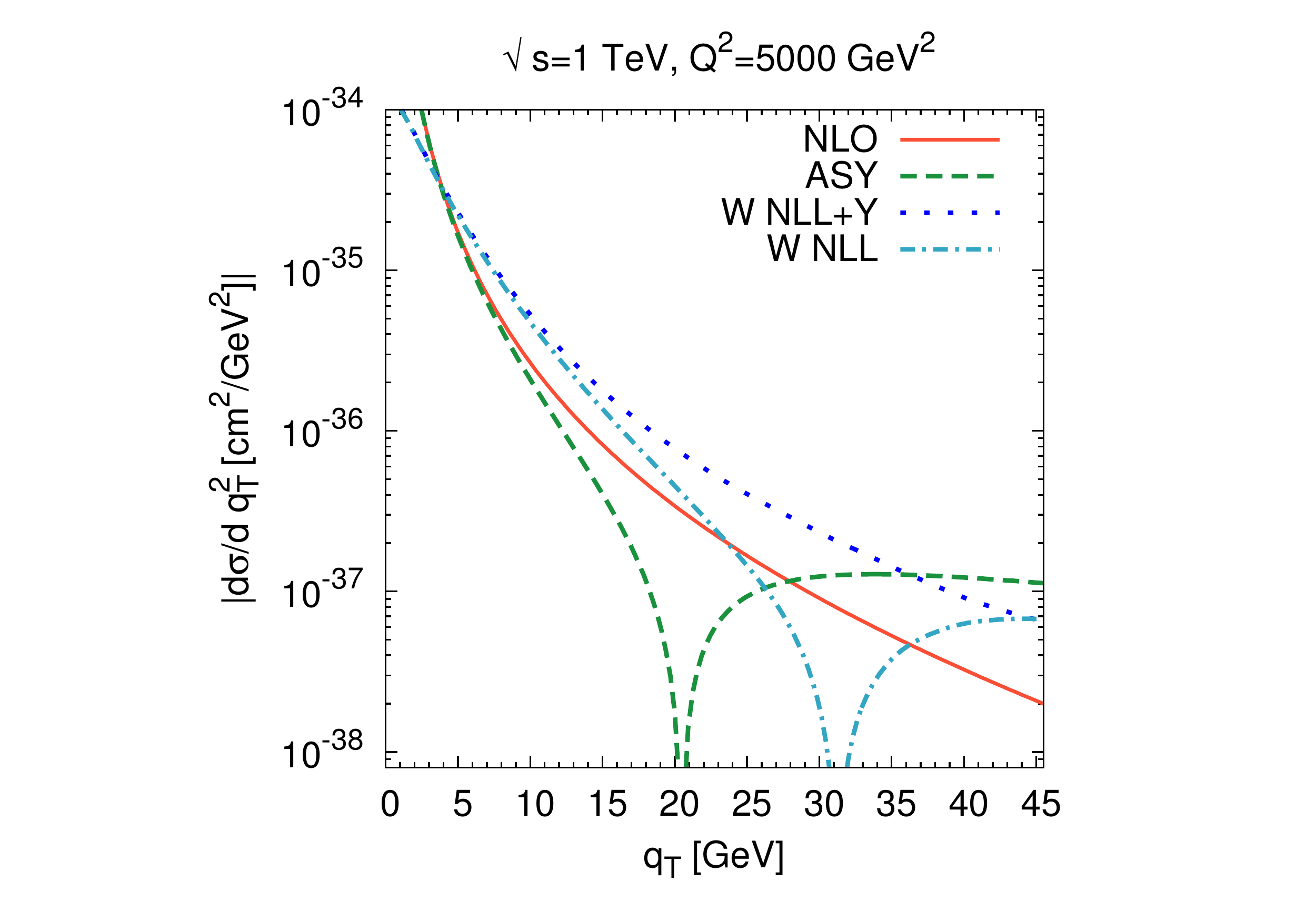} \hspace*{-2.5cm}
\includegraphics[width=7.4cm]{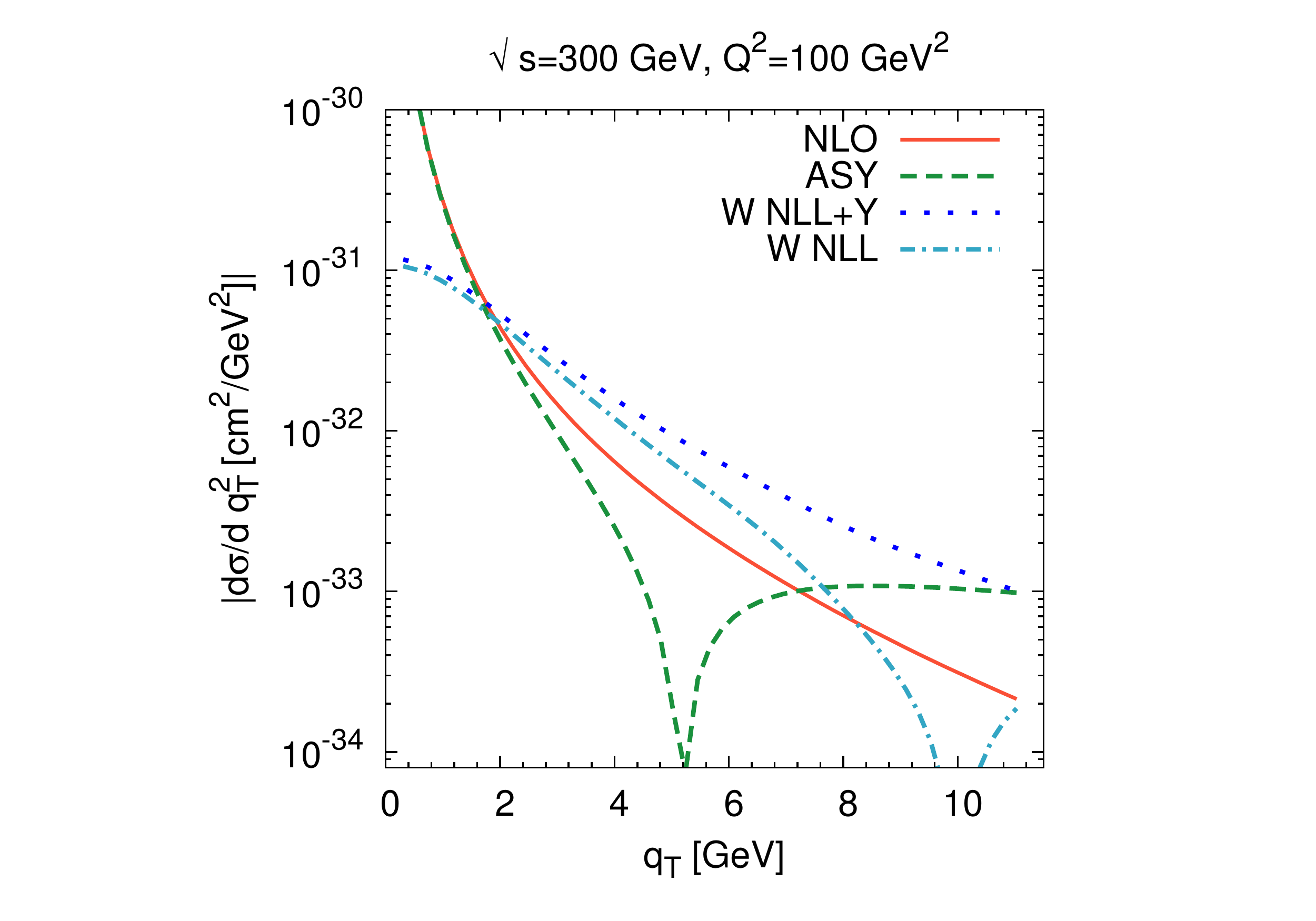} \hspace*{-2.5cm}
\includegraphics[width=7.4cm]{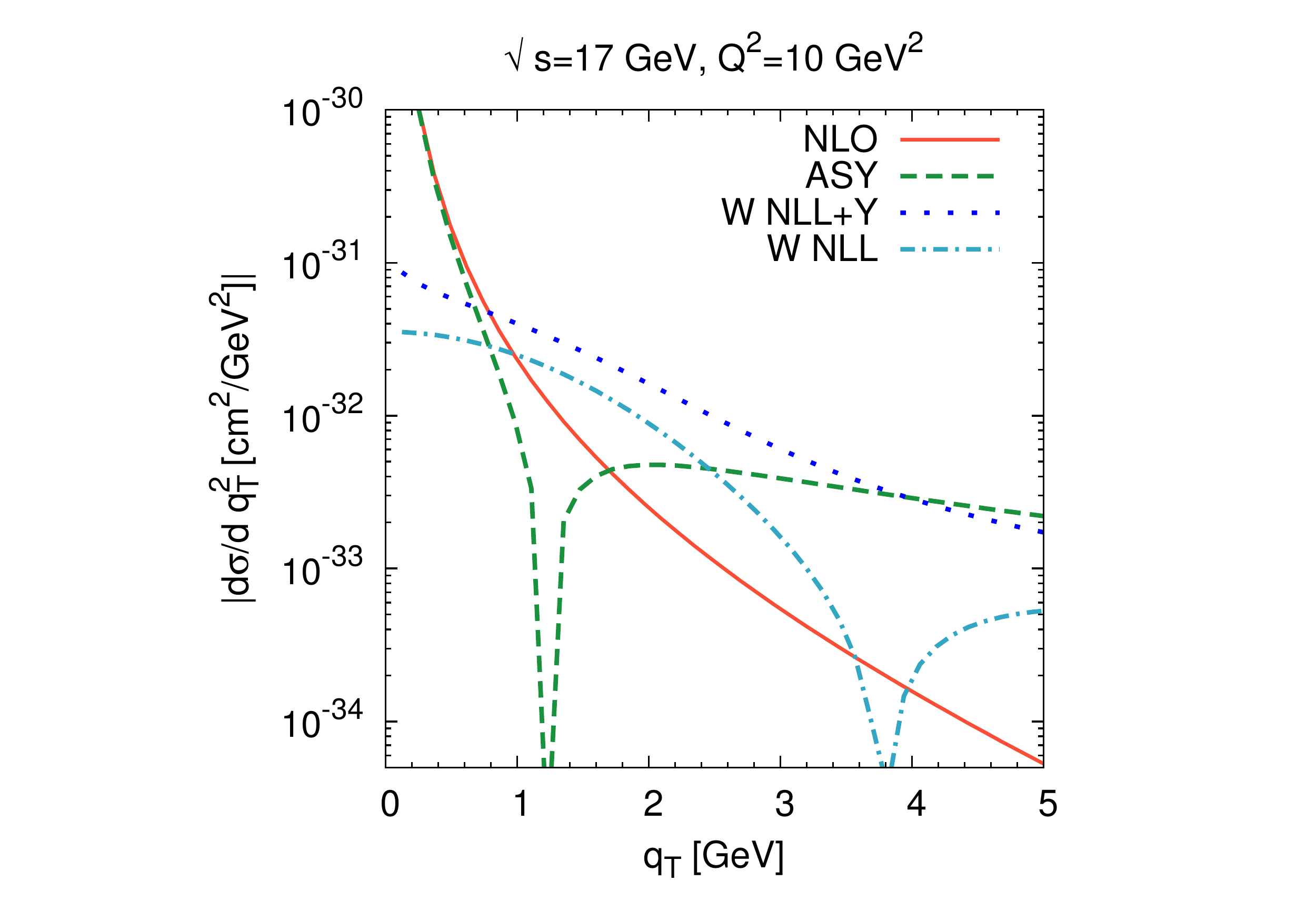}  }
\vspace*{-8pt}
\caption{
$d\sigma^{NLO}$, $d\sigma^{ASY}$, 
$W^{NLL}$ and the sum $W^{NLL}+Y$ (see Eq.~\eqref{eq:yterm}), corresponding to the three 
different SIDIS kinematical configurations defined in Fig.~\ref{f1}.
Here $b_{max}=1.0$ GeV$^{-1}$, $g_1=0.3$ GeV$^2$,  $g_{1f}=0.1$ GeV$^2$, $g_2=0$ GeV$^2$.
 \label{f6}}
\end{figure}

In Fig.~\ref{f6} we show, in the three SIDIS configurations considered above, 
the NLO cross section $d\sigma^{NLO}$ (solid, red line), the asymptotic cross section $d\sigma^{ASY}$ (dashed, green line) 
and the NLL resummed cross section $W^{NLL}$ (dot-dashed, cyan line).
The dotted blue line represents the sum $(W^{NLL}+Y)$, according to  Eq.~\eqref{SIDIS-CSS1}.
%

Clearly, in none of the kinematical configurations considered, $W^{NLL}$ matches $d\sigma^{ASY}$, 
they both change sign at very different values of $q_T$.
Moreover, the $Y$ factor can be very large compared to $W^{NLL}$. 
Consequently, the total cross section $W^{NLL}+Y$ (dotted, blue line) never matches the fixed order cross section 
$d\sigma^{NLO}$ (solid, red line).
At low and intermediate energies, the main source of the matching failure is represented by the non-perturbative contribution 
to the Sudakov factor.
As we showed in Section~\ref{sec:NP},  the resummed term $W$ of the cross section is totally dominated by the non-perturbative input, 
even at large $q_T$.
Notice that, in the kinematical configurations of the COMPASS experiment, the matching cannot be achieved simply by 
adding higher order corrections to the perturbative calculation of the $Y$ term, as proposed in Ref.~\cite{Su:2014wpa}, 
as 
$W^{NLL}$ is heavily dependent on the non-perturbative input. 

 Interestingly, the cross section does not match the NLO result even at 
 the highest energies considered, $\sqrt{s} =1$ TeV and $Q^2 = 5000$ GeV$^2$: further comments will be addressed 
 in the following subsection.

\subsection{Matching with the inclusion of non-perturbative contributions \label{sec:match2}}

As discussed above, the mismatch between $W^{NLL}$ and $d\sigma^{ASY}$ at $q_T\sim Q$ is mainly due to 
the non-perturbative content of the cross section, which turns out to be non-negligible, at least at 
low and intermediate energies.
To try solving this problem one could experiment different and more elaborate matching prescriptions, 
which somehow take into account the non-perturbative contributions to the total cross section.  
In alternative to $d\sigma^{total}= W^{NLL} + Y$, Eq.~\eqref{eq:yterm}, one could require, for instance, that 
in a region of sizable $q_T$
\be
d\sigma^{total
} = W^{NLL} - W^{FXO} + d\sigma^{NLO}\,,
\label{match2}
\ee
where $W^{FXO}$ is the  NLL resummed cross section approximated at first order in $\alpha_s$,
with a first order expansion of the Sudakov exponential, $\exp[S_{pert}(b_*)]$.   
The result for the Fixed Order (FXO) expansion of $W^{S\!I\!D\!I\!S}$ is presented in Eq.~(\ref{eq:FXO}) 
of the Appendix. 
Notice that our FXO expansion differs from that proposed in Ref.~\cite{Koike:2006fn}, 
where the scale of $\alpha_s$ used for the perturbative expansion 
of the cross section is taken to be equal to the factorization scale. In our computation this scale is 
simply $\mu_b$: with our choice, the FXO result is closer to that obtained by using the power 
counting of $W^{NLL}$, see Section~\ref{resummation}. Instead, the result of Ref.~\cite{Koike:2006fn} 
is more in line with the fixed order $\alpha_s$ expansion performed in the calculation of $d\sigma^{NLO}$. 
In principle, the two approaches should be the same when terms proportional to $\log(Q^2/\mu_b^2)$ 
are small and both coincide up to $\alpha_s^2$ corrections.
%
 \begin{figure}[t]
 \centerline{
 \includegraphics[width=7.4cm]{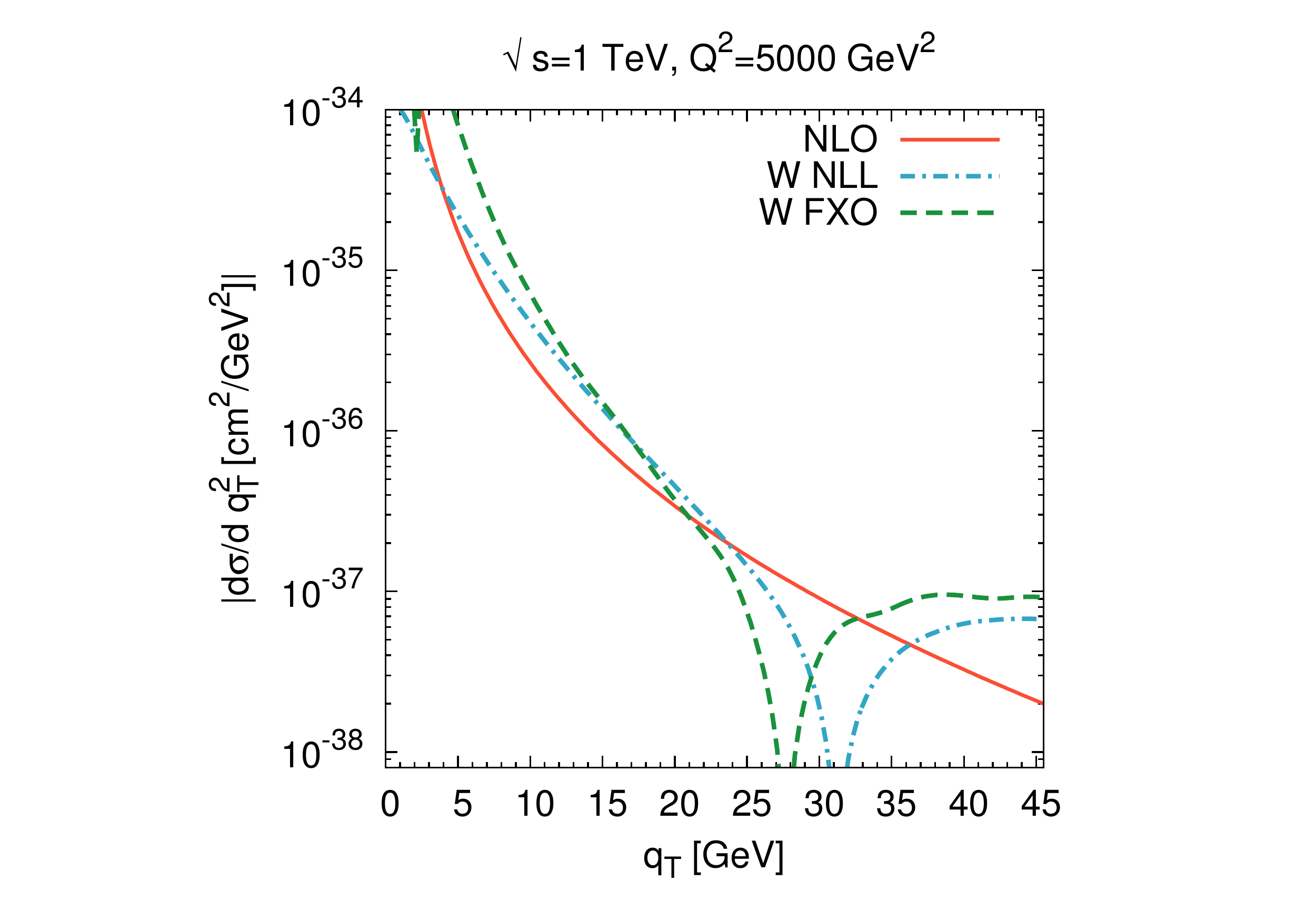}\hspace*{-2.5cm}
 \includegraphics[width=7.4cm]{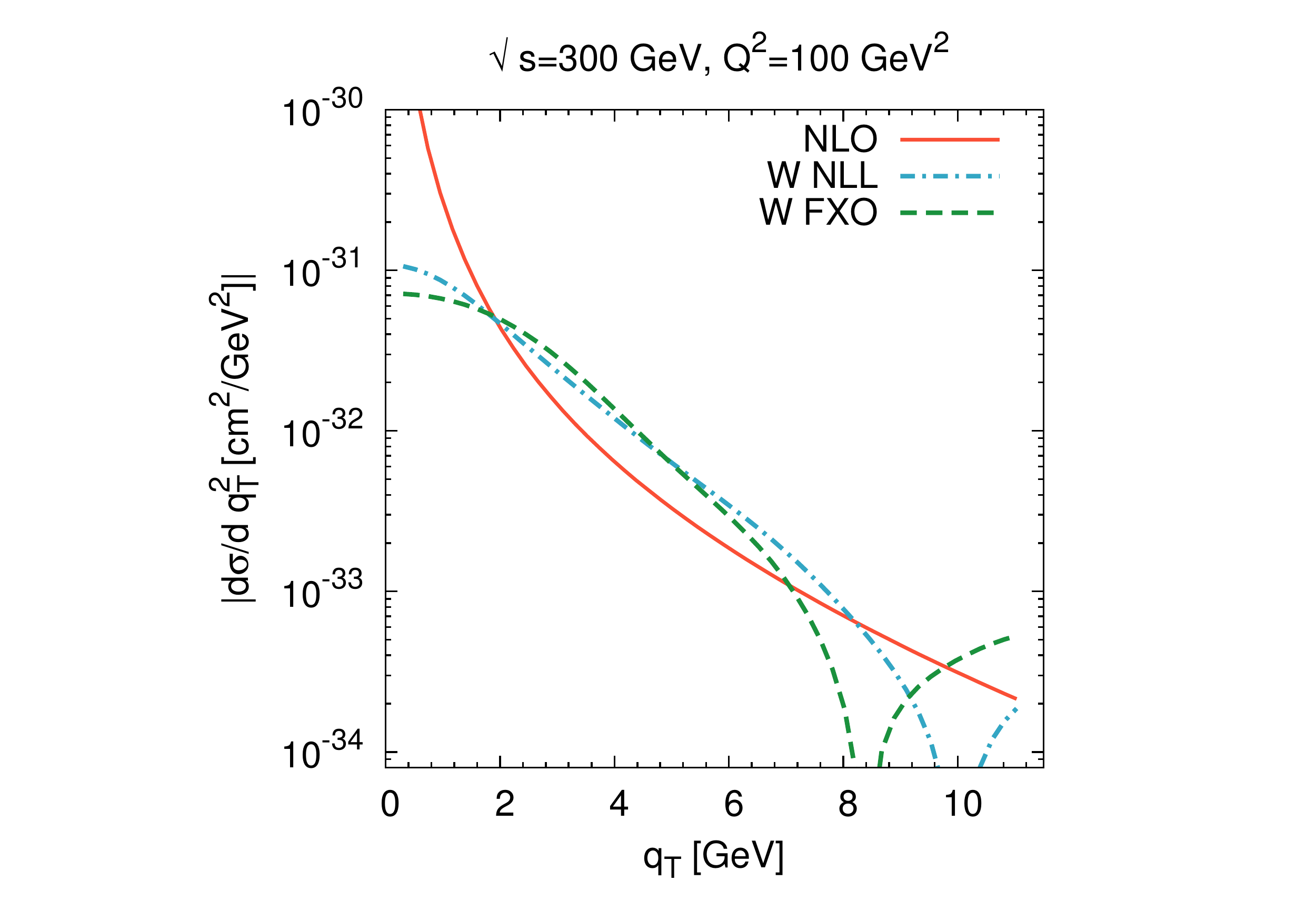}\hspace*{-2.5cm}
 \includegraphics[width=7.4cm]{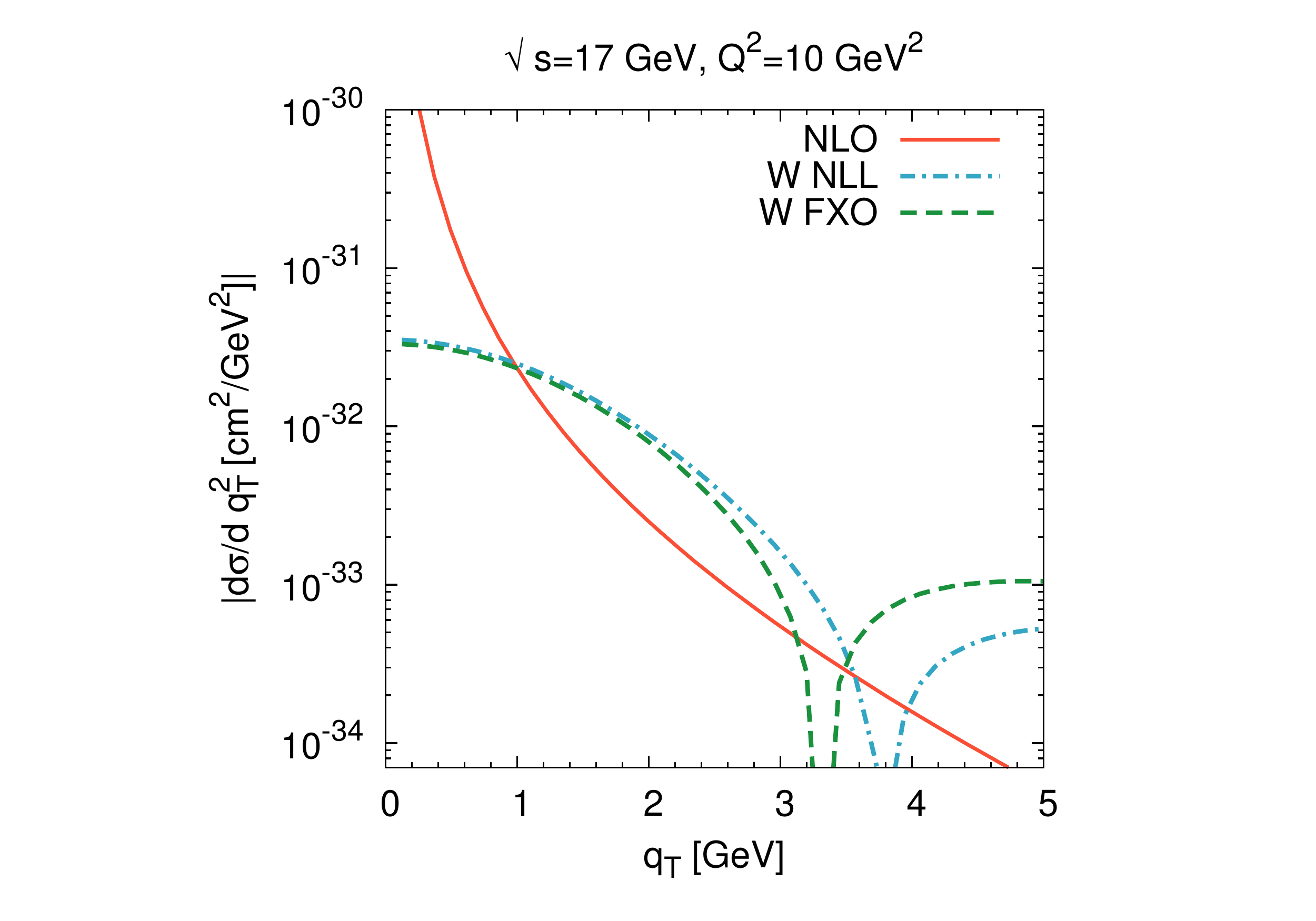}}
 \vspace*{-8pt}
 \caption{
 $d\sigma^{NLO}$, 
$W^{NLL}$ and $W^{FXO}$ (see Eq.~\eqref{match2}), corresponding to three 
different SIDIS kinematical configurations.
Here $b_{max}=1.0$ GeV$^{-1}$, $g_1=0.3$ GeV$^2$,  $g_{1f}=0.1$ GeV$^2$, $g_2=0$ GeV$^2$.
 \label{f7}}
 \end{figure}

As mentioned above, we build $W^{FXO}$ so that it contains 
the same non-perturbative Sudakov, $S_{NP}$, we assign to $W^{NLL}$: 
therefore we might expect to find a region in which $W^{FXO} \simeq W^{NLL}$, allowing to match the SIDIS 
cross section $d\sigma = W^{NLL} - W^{FXO} + d\sigma^{NLO}$ to the purely perturbative 
cross section $d\sigma^{NLO}$.  

On the other hand, in the absence of non-perturbative content inside $W^{FXO}$ and $W^{NLL}$, and in the perturbative limit, when  $\exp[S_{pert}]$ 
can be approximated by $1+S_{pert}$, with $S_{pert}$ expanded at first order in $\alpha_s$, 
one can show that $W^{FXO} \to d\sigma^{ASY}$  so that, in this region~\cite{Kawamura:2007gh,Nadolsky:2001sf}
\be
d\sigma^{total} = W^{NLL} - W^{FXO} + d\sigma^{NLO} \to  W^{NLL} - d\sigma^{ASY} + d\sigma^{NLO}= W^{NLL} + Y\,.
\ee
In this limit this prescription is equivalent to the Y-term matching prescription of Eq.~(\ref{eq:yterm}).

Fig.~\ref{f7} shows $d\sigma^{NLO}$ (solid, red line), $W^{NLL}$ (dash-dotted, cyan line) and $W^{FXO}$ (dashed, green line)  
for the same three kinematical configurations 
considered in the previous plots. 
At 1 TeV and in the HERA kinematical configuration, there is some region in which $W^{FXO}$  and $W^{NLL}$ are crossing.
However, this does not happen at $q_T \sim Q$, where one would expect to match to $d\sigma^{NLO}$. 
Contrary to our expectations, we do not find a region in which $W^{NLL}$ coincides asymptotically to its expansion $W^{FXO}$, up to numerical 
precision and higher order corrections. Therefore, no smooth and continuous matching can be performed.
For the COMPASS-like experiment, where the non-perturbative regime basically dominates the whole cross section, the $W^{FXO}$ and $W^{NLL}$ 
curves never cross, see the right panel of Fig.~\ref{f7}. Therefore no matching whatsoever is possible.

\begin{figure}[t]
\centerline{
\includegraphics[width=7.4cm]{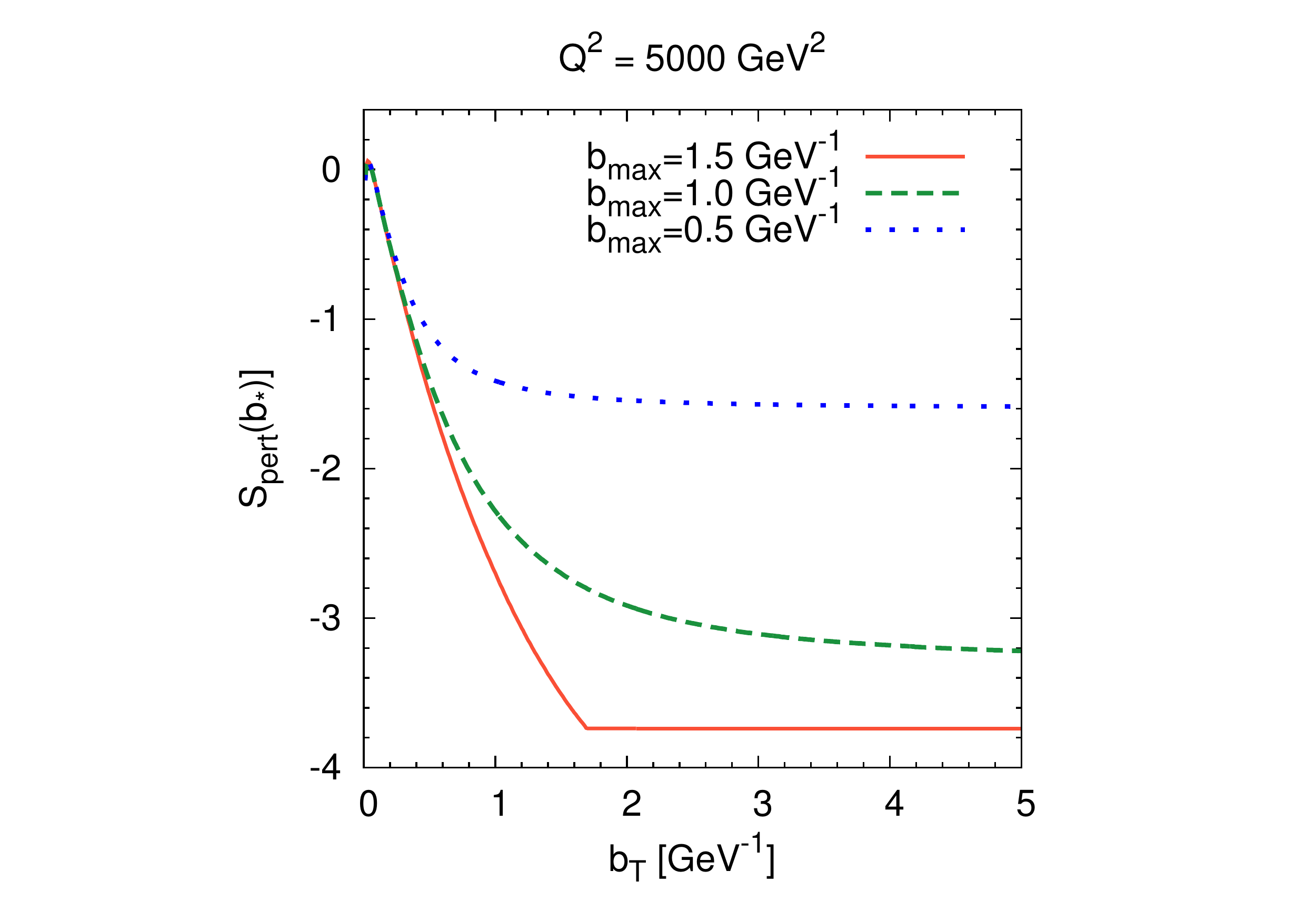}    \hspace*{-2.5cm}
\includegraphics[width=7.4cm]{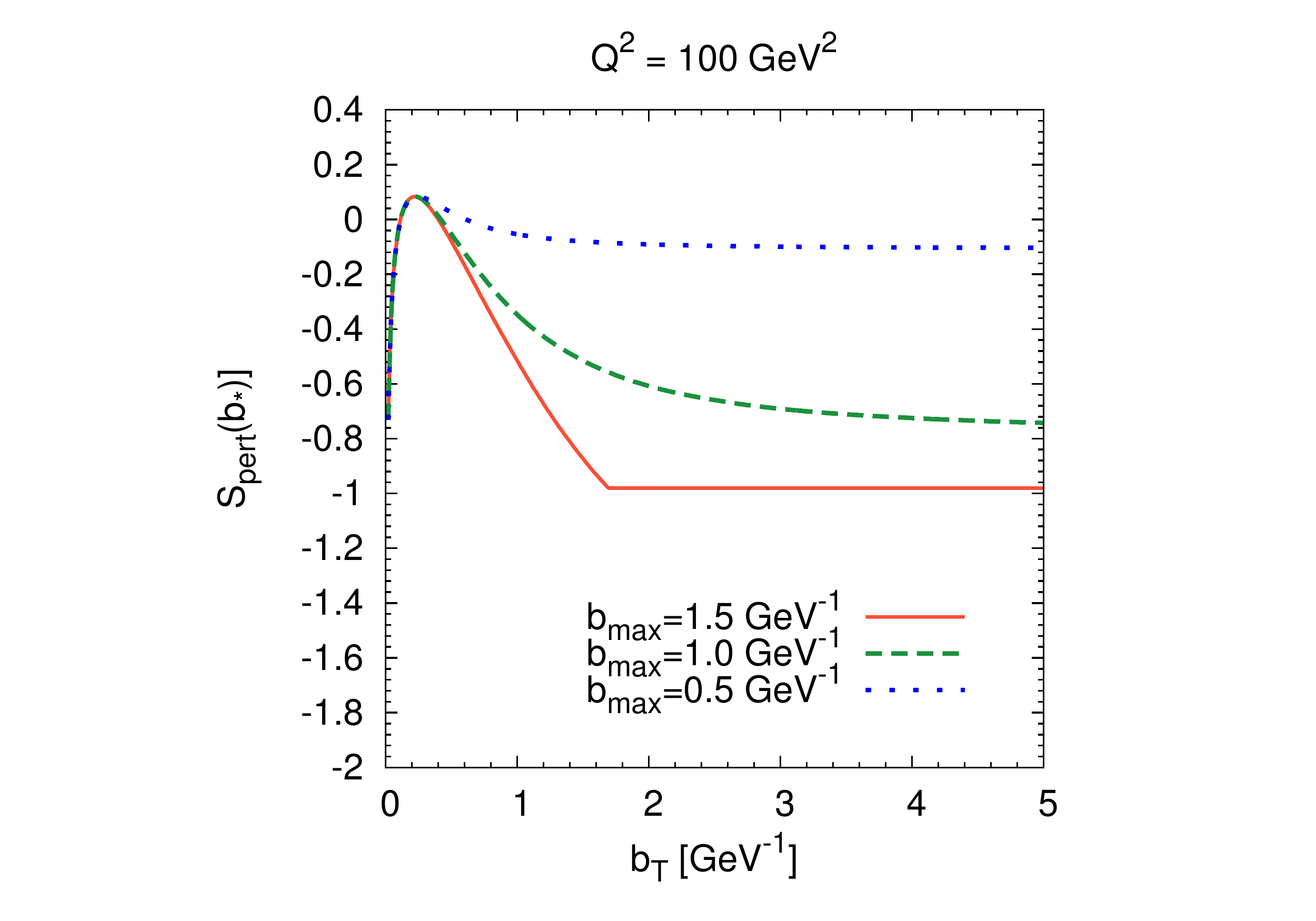} \hspace*{-2.5cm}
\includegraphics[width=7.4cm]{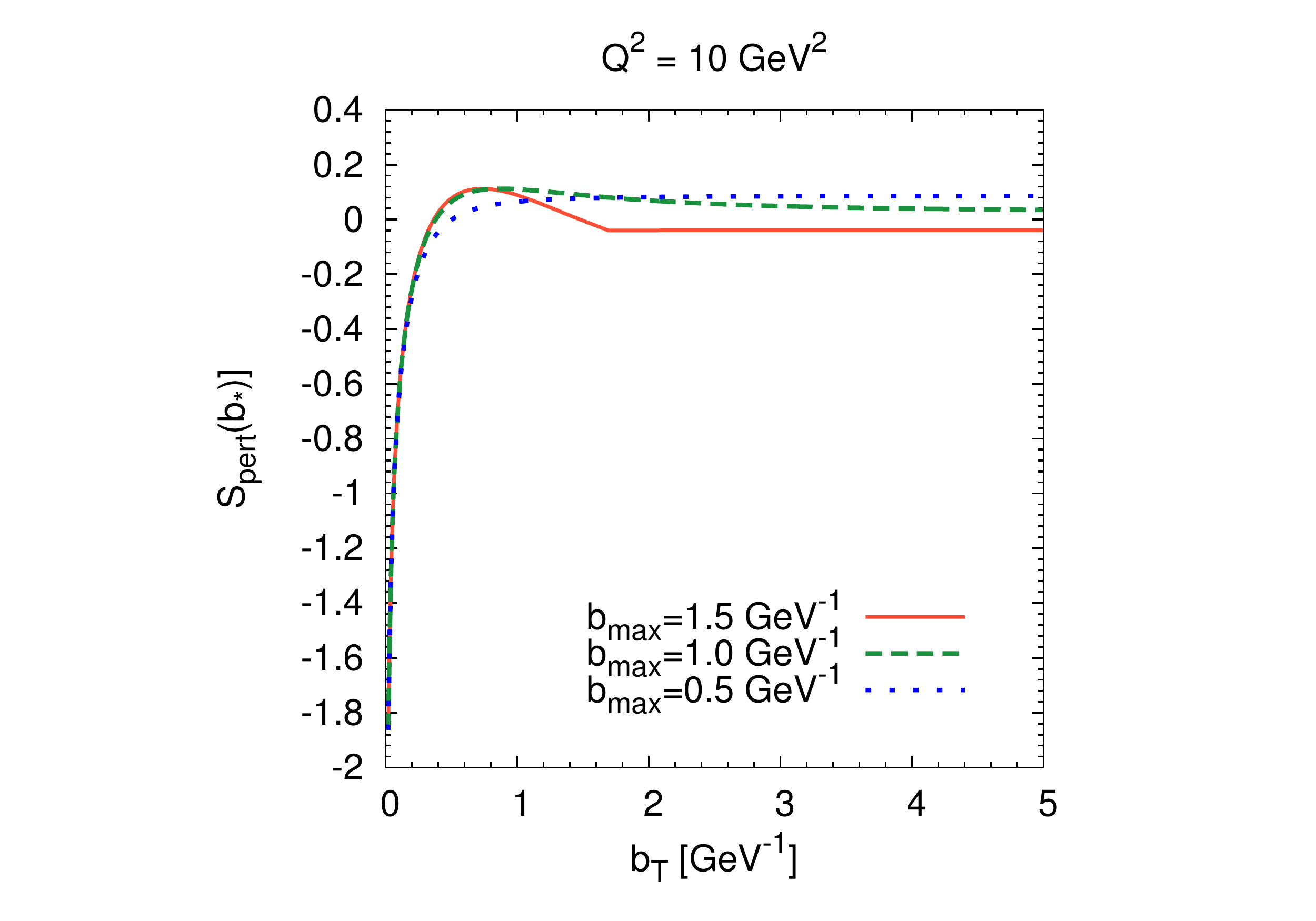} 
} 
\vspace*{-8pt}
\caption{ 
The perturbative Sudakov factor $S_{pert}(b_*)$. On the left panel $Q^2=5000$ GeV$^2$, 
on the central panel $Q^2=100$ GeV$^2$, and on the right panel
$Q^2=10$ GeV$^2$. We consider three values of $b_{max}$: 
$1.5$ GeV$^{-1}$(solid line), $1$ GeV$^{-1}$(dashed line), $0.5$ GeV$^{-1}$ (dotted line).
\label{f8}}
\end{figure}

Let's summarize: in the previous Section we have shown that the Y-term matching prescription does not work, even at high energies. 
Here we adopted a different prescription, which takes into account the non-perturbative Sudakov contribution. 
Also in this case we find that the matching fails. 

To understand the reason of this failure, we shall investigate 
the behaviour of the Sudakov factor in more detail.
As explained in Appendix~\ref{A}, the fixed order expansion of the W-term, $W^{FXO}$, 
is computed by expanding the perturbative Sudakov exponential to first order in  $S_{pert}$, $\exp[S_{pert}] \sim 1 + S_{pert}$, 
and considering the whole $W$ to first order in $\alpha_s$. 
Indeed, this expansion holds only when successive powers of $\alpha_s$ are small, when the logarithmic terms are small and 
consequently when $S_{pert}$ itself is small.

Fig.~\ref{f8} shows that the Sudakov factor $S_{pert}$ is small only in a limited region of $b_T$ depending 
on the kinematical details of the SIDIS process (at 1 TeV this region is very narrow).  
Instead, at very small and large $b_T$, the Sudakov factor $S_{pert}$ is large. Notice also that, at large $b_T$, 
its size strongly depends on the choice of $b_{max}$.

\begin{figure}[t]
 \centerline{
 \includegraphics[width=7.4cm]{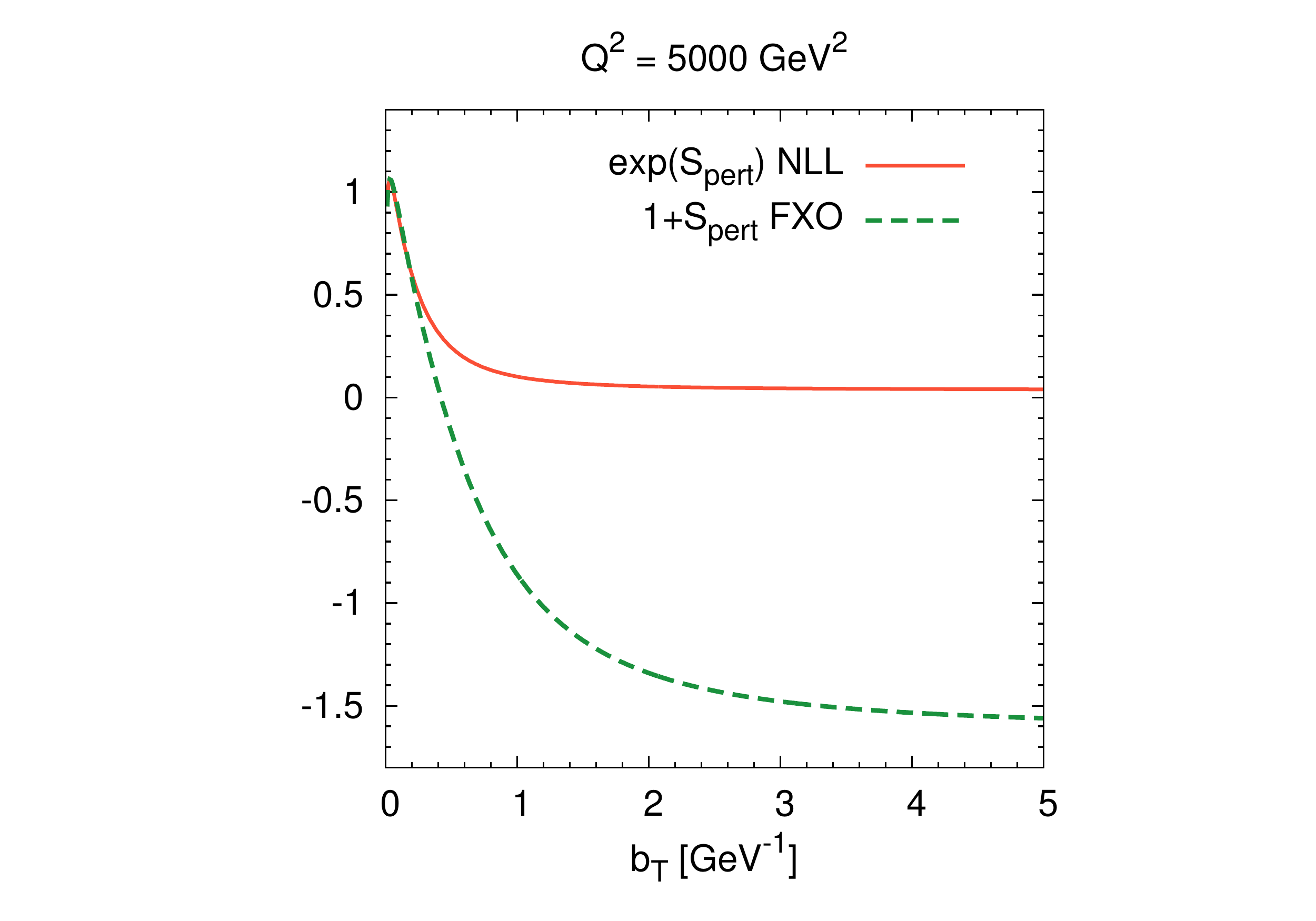}\hspace*{-2.5cm}
 \includegraphics[width=7.4cm]{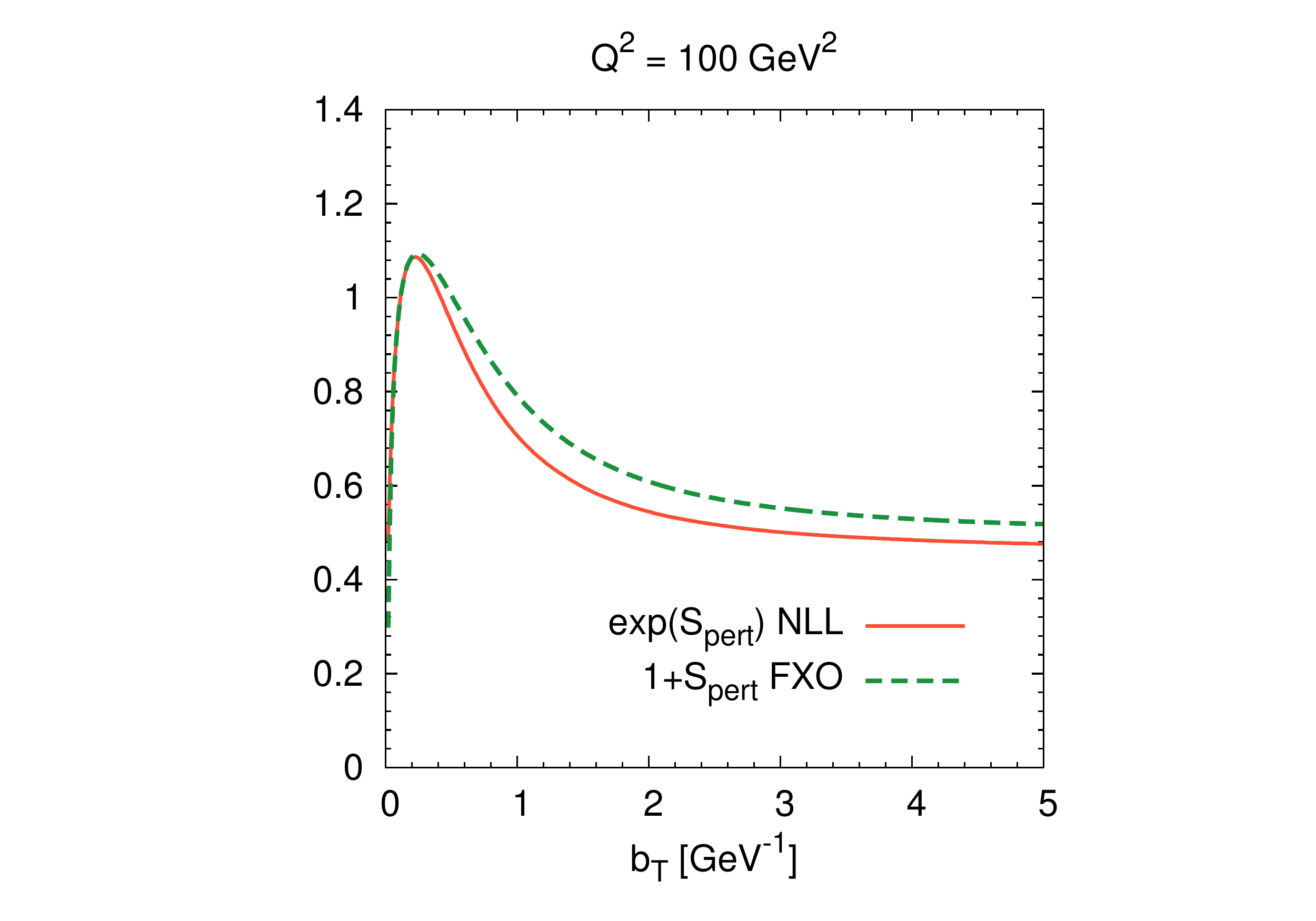}\hspace*{-2.5cm}
 \includegraphics[width=7.4cm]{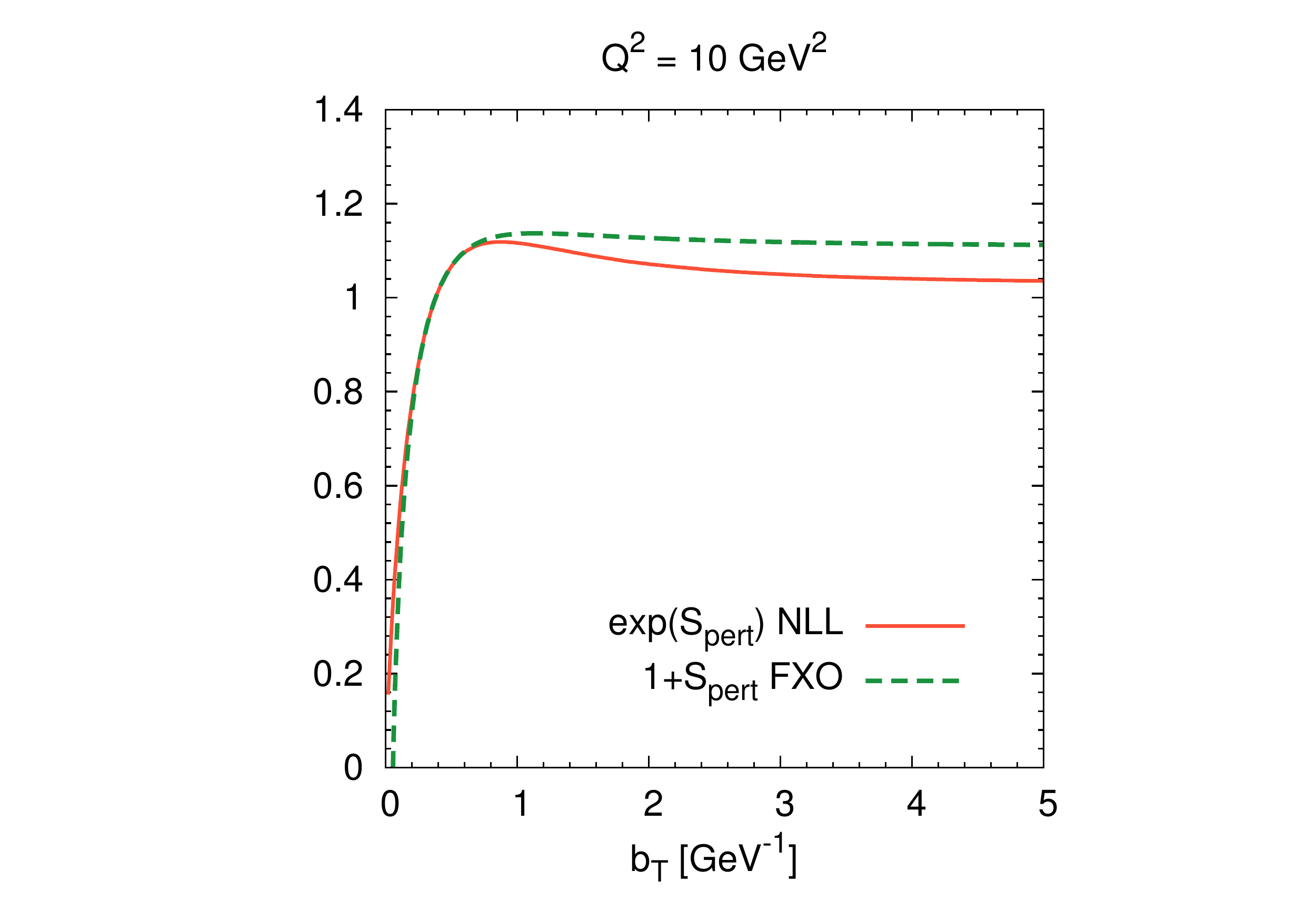}}
 \vspace*{-8pt}
 \caption{The perturbative Sudakov exponential $\exp[S_{pert}^{NLL}]$ (solid line) 
 and its expansion $1 + S_{pert}^{FXO}$ (dashed line). 
On the left panel $Q^2=5000$ GeV$^2$, 
on the central panel $Q^2=100$ GeV$^2$, and on the right panel
$Q^2=10$ GeV$^2$. We fix $b_{max}=1$ GeV$^{-1}$.
 \label{f9}}
 \end{figure}

In Fig.~\ref{f9} we plot $\exp[S_{pert}^{NLL}]$ and its expansion $1 + S_{pert}^{FXO}$. 
Notice that two steps are involved in this expansion:  
\be
\exp[S_{pert}^{NLL}] \to \exp[S_{pert}^{FXO}] \to  1 + S_{pert}^{FXO} \,.
\ee
The differences between $\exp[S_{pert}^{NLL}]$ and $1 + S_{pert}^{FXO}$ 
are therefore due to two reasons: $S_{pert}^{NLL}$ and $S_{pert}^{FXO}$ are different and, in general, 
they are small only in a limited range of $b_T$.
As one can see in Fig.~\ref{f9}, these differences occur  
in both the small and the large $b_T$ regions.

The authors of Refs.~\cite{Ellis:1997ii,Frixione:1998dw} pointed out that  the 
Sudakov factor~\cite{Altarelli:1984pt} vanishes at $b_T=0$ in the exact first order calculation.
To restore this behaviour of the CSS Sudakov factor, prescriptions exist in the literature which ensure  
$S_{pert} \to 0$ at $b_T\to 0$. 
After integration, the Sudakov form factor can be written as a function of $\log(Q^2/\mu_b^2)=\log(Q^2 b_T^2/C_1^2)$, 
which become large and negative at $b_T\to 0$.
A suggested prescription to avoid this problem, consists in replacing 
\be
\log(Q^2/\mu_b^2) \to \log(1+Q^2/\mu_b^2)\,,
\label{w-prescr}
\ee
see for example Ref.~\cite{Frixione:1998dw, Koike:2006fn}.

\begin{figure}[t]
 \centerline{
 \includegraphics[width=7.4cm]{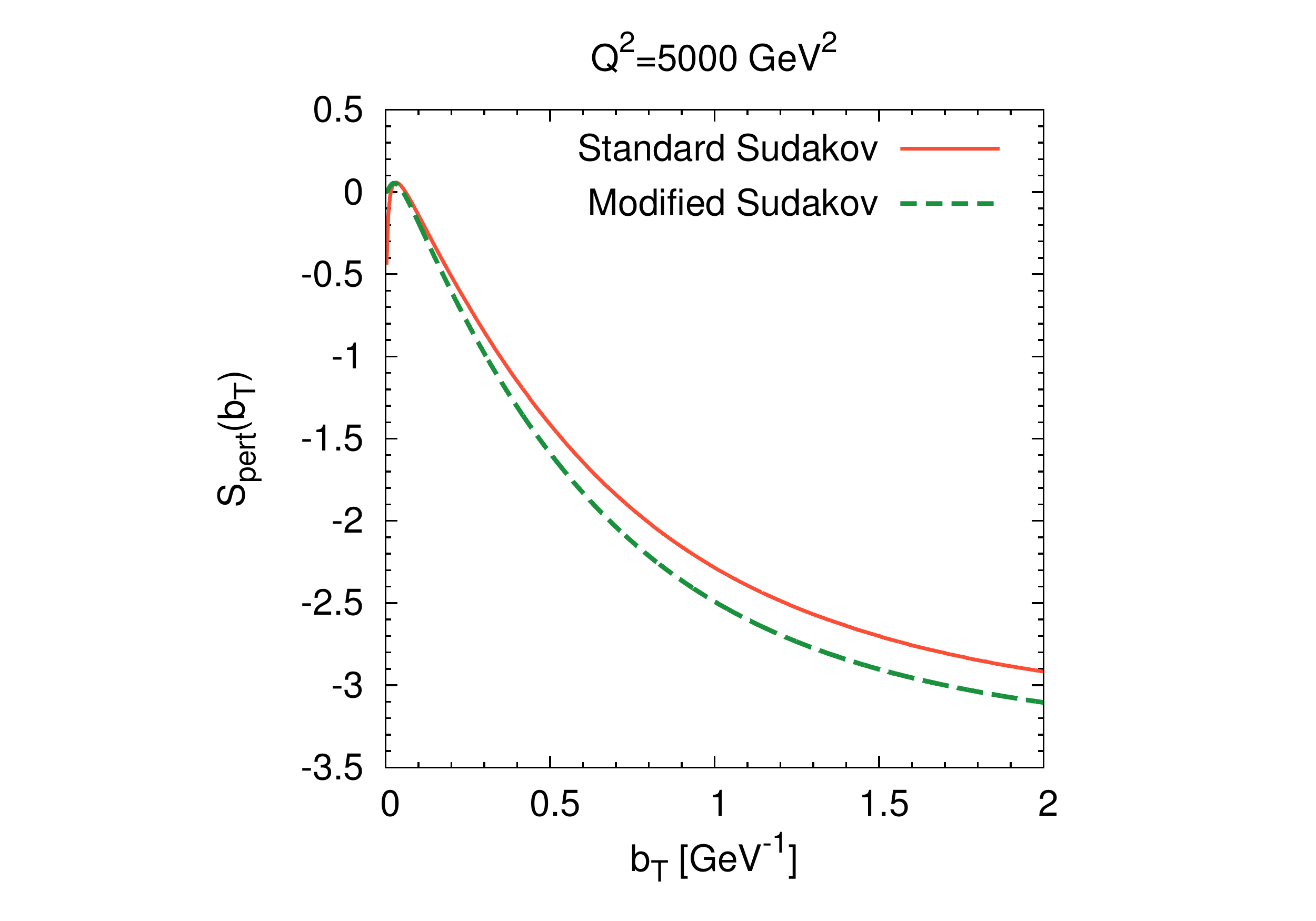}\hspace*{-2.5cm}
 \includegraphics[width=7.4cm]{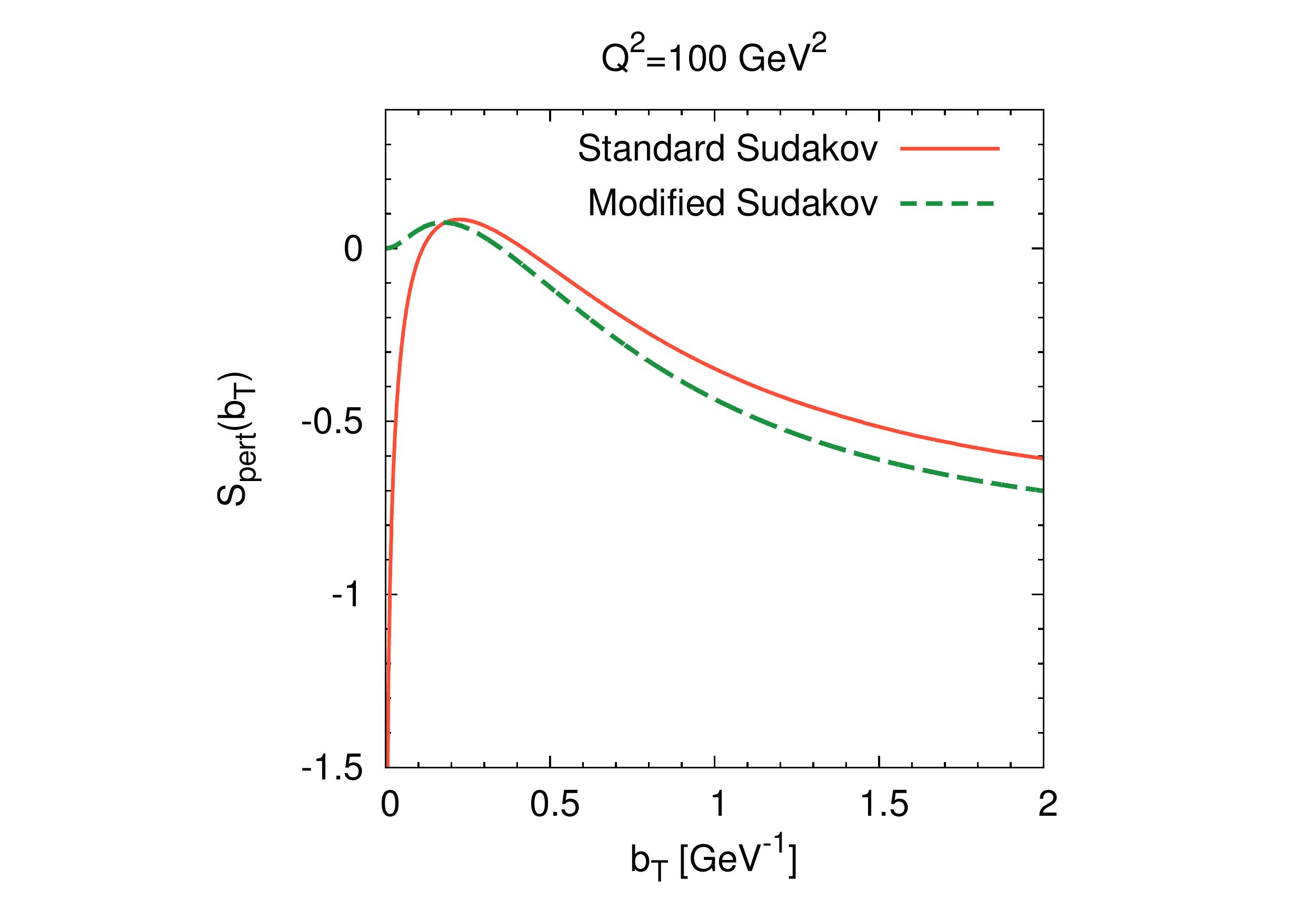}\hspace*{-2.5cm}
 \includegraphics[width=7.4cm]{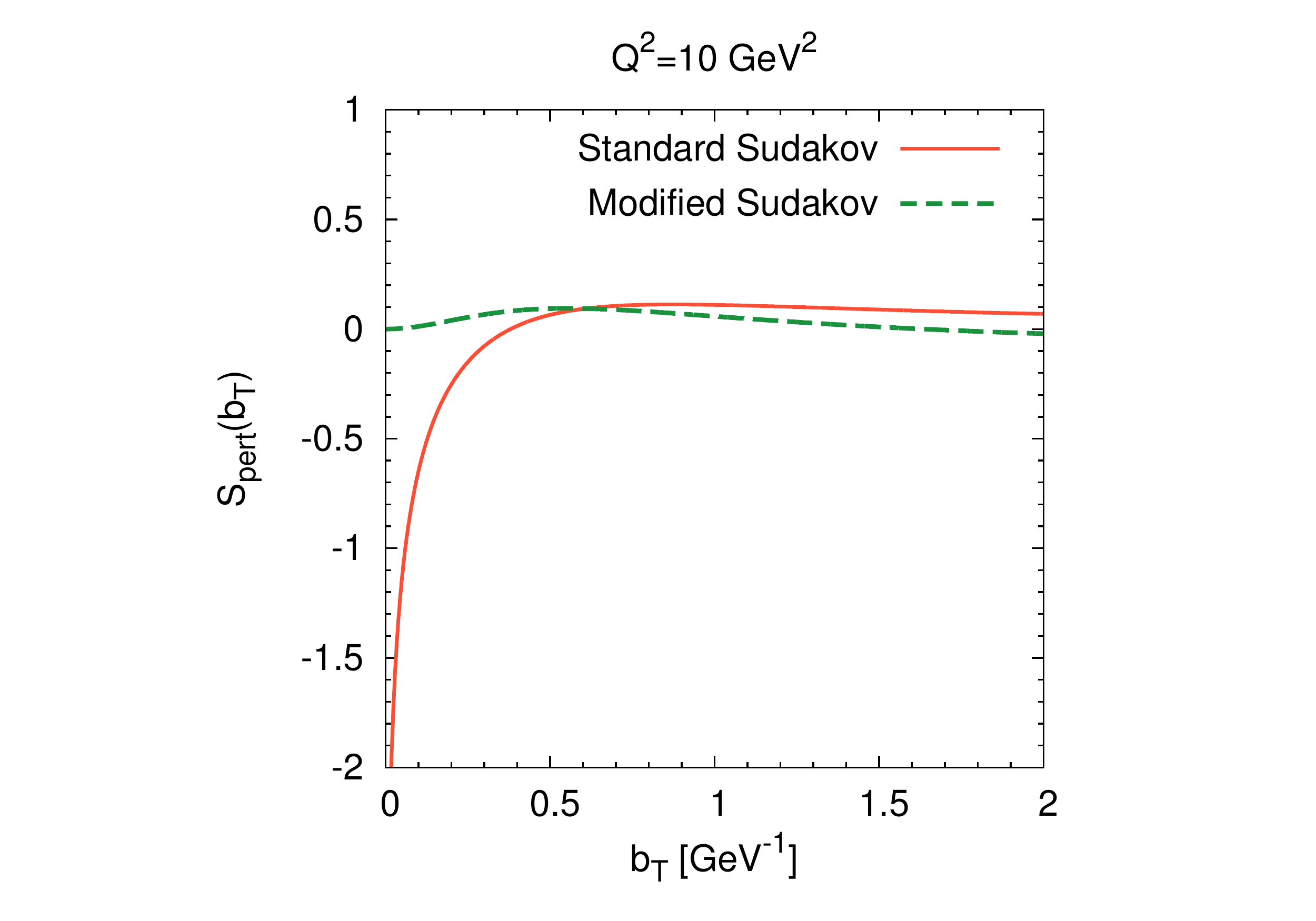}}
 \vspace*{-8pt}
 \caption{Sudakov factor as given by Eq.~\eqref{S} (solid line), and its modified form given in  
Eqs.~(44)-(47) of Ref.~\cite{Koike:2006fn} (dashed line), for three different values of $Q^2$.
 \label{f10}}
 \end{figure}

The effect of this recipe can bee visualized in Fig.~\ref{f10}, where the standard, Eq.~\eqref{S}, and modified, 
Eqs.~(44)-(47) of Ref.~\cite{Koike:2006fn},  forms of the Sudakov factor are compared, 
for three different kinematical configurations. Clearly, the plots show that this prescription has a much stronger effect 
at small $Q^2$ than at large $Q^2$: the failure of the matching prescription at 1 TeV is therefore not solved, 
however a better result  might be achieved for the smaller energy configurations (HERA and COMPASS). 

One can see from Fig.~\ref{f8}-\ref{f10} that 
the perturbative Sudakov factor 
 $S_{pert}(b_*)$ in some regions of $b_T$ is positive, i.e. $\exp[S_{pert}(b_*)]>1$ allowing for an unphysical
Sudakov  enhancement. 
In particular in COMPASS-like kinematics, this enhancement dominates over almost all the $b_T$ range while 
at higher energies its relevance is limited.
This is a signal of the inadequacy of the resummation approaches at such low energies.
%

We have checked that, even adopting the prescription of Eq.~\eqref{w-prescr}, for the 1 TeV kinematical configuration 
the matching cannot be performed. In fact, the impact of this prescription is rather limited in this case. 
The failure of the matching is likely due to the fact that the perturbative expansion of the Sudakov factor  
breaks down at a very early stage in $b_T$, see the top-left panel of Fig.~\ref{f8} and the left panel of Fig.~\ref{f9}.

The HERA configuration deserves a dedicated discussion. 
We can observe that, adopting the method of Eq.~\eqref{w-prescr}, the Sudakov exponential can be quite successfully expanded as  
$\exp[S_{pert}] \sim 1 + S_{pert}$ over the whole $b_T$ range, see the central panels of Figs.~\ref{f8} and~\ref{f9}.

\begin{figure}[t]
 \centerline{
 \includegraphics[width=10.cm]{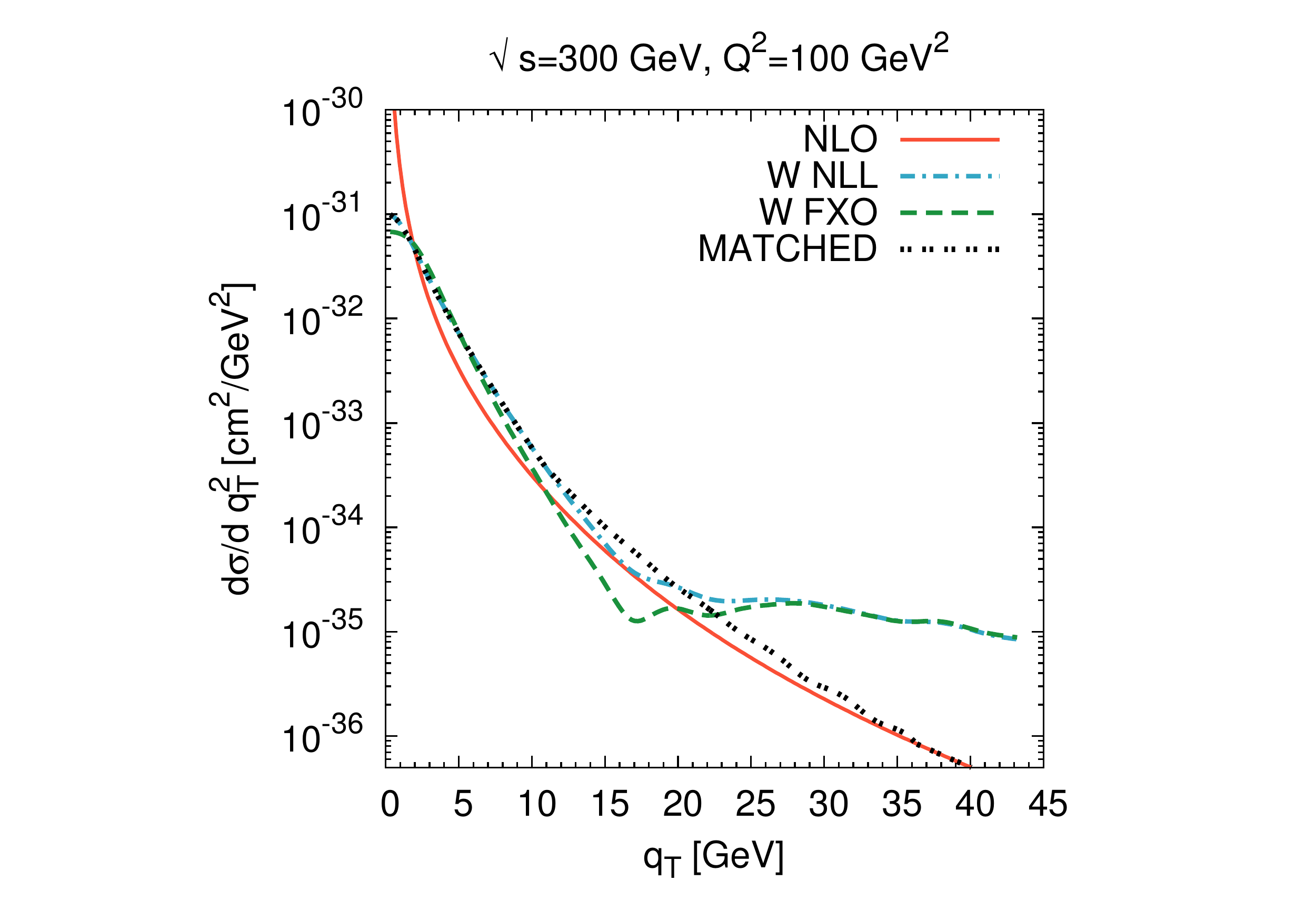}
 }
 \vspace*{-8pt}
 \caption{ $d\sigma^{NLO}$, 
$W^{NLL}$ and $W^{FXO}$ (see Eq.~\eqref{match2}), corresponding to the HERA-like kinematical configurations.
Here the low $b_T$ behaviour of the Sudakov factor has been corrected using Eq.~\eqref{w-prescr}. 
The double-dotted black line represents the matched cross section, as described in the text. 
We fix $b_{max}=1.0$ GeV$^{-1}$, $g_1=0.3$ GeV$^2$,  $g_{1f}=0.1$ GeV$^2$, $g_2=0$ GeV$^2$. 
Notice that all contributions are positive in this case.
 \label{f11}}
 \end{figure}

In this case, in fact, a region where $W^{NLL}$ and $W^{FXO}$ approximately match actually exists, as shown in Fig.~\ref{f11}.
This means that here, for this particular kinematical configuration, the perturbative expansion works and all the conditions 
required for the matching seem to be approximately fulfilled. 
In order to achieve a fully matched cross section, one also needs to know where to start using $W^{NLL} - W^{FXO} + d\sigma^{NLO}$  instead of $W^{NLL}$: 
this can happen in the region where $W^{FXO} \sim d\sigma^{NLO}$.
Ideally, in the absence of any non-perturbative contributions,  $W^{FXO} \sim d\sigma^{ASY}$ at small $q_T$, where $d\sigma^{NLO} \sim d\sigma^{ASY}$, 
allowing for a region of successful matching.
However, since $W^{FXO}$ is affected by a sizable non-perturbative content, it turns out to be different from $d\sigma^{ASY}$ and therefore different 
from $d\sigma^{NLO}$ at small $q_T$.
In this case, there will be at most one crossing point between the $W^{FXO}$ and the $d\sigma^{NLO}$ curves, which does not provide a smooth matching.

Indeed, one should remember  that all these contributions
are computed within theoretical errors due, for instance, to the choice of renormalization scale and to the truncation of the 
perturbative series. Consequently, one could think that a smooth matching 
could be achieved within the corresponding error bands, 
rather than on individual points of the single curves, through an interpolating function.

%
%

 \section{Conclusions and outlook\label{Conclusions}}

Soft and collinear gluon resummation in the impact parameter $b_T$ space is a very powerful tool. 
However, its successful implementation is affected by a number of practical difficulties: the strong influence of the 
kinematical details of the SIDIS process, 
the possible dependence of the parameters used to model the non-perturbative content of the SIDIS cross section, 
the complications introduced by having to perform phenomenological studies in the $b_T$ space, 
where the direct connection to the conjugate $q_T$ space is lost.

Indeed, matching prescriptions have to be applied to achieve a reliable description of the SIDIS process
over the full $q_T$  range, going smoothly from the region of applicability of resummation, or equivalently of the TMD description,  to the region of 
applicability of perturbative QCD.

In any resummation scheme, one needs to take care of the non-perturbative content. Here we adopt the so-called $b_*$ 
prescription in order to cure the problem of the Landau pole in the perturbative expansion, complementing it with the 
introduction of a properly defined non-perturbative function. In Subsections~\ref{sec:NP} and~\ref{sec:bmax} we studied 
the dependence of our results on this non-perturbative contribution and on the details of the  $b_*$ prescription, 
i.e. on $b_{max}$. We found that some kinematical configurations, similar to those of COMPASS or HERMES experiments 
for example, are completely dominated by these features.
Therefore, in Subsection~\ref{sec:match} we concluded that no matching can be achieved exploiting the Y-term which, being  
calculated in perturbative QCD, does not include any non-perturbative content.

To address this problem, we adopted a different matching prescription, Eq.~\eqref{match2}, which takes into account 
(and include) all details of the non-perturbative behaviour. However, this method still presents several difficulties and 
remains largely unsatisfactory.  
In order to find the origin of these difficulties, we studied in detail the $b_T$ behaviour of the perturbative Sudakov 
factor, in three different kinematical configurations. 
We found that in a COMPASS-like  kinematical configuration the perturbative Sudakov exponential is larger than one, 
i.e. unphysical, over most of the $b_T$ range. 
Therefore any resummation scheme would be inadequate in this case, and hardly applicable.
Instead, for the other two kinematical configurations analyzed, $\exp[S_{pert}]>1$ only on a limited range of $b_T$, 
thus not affecting the results in the $q_T$ space. Nevertheless, even in these cases, the matching prescription of 
Eq.~\eqref{match2} does not work
as the expansion $\exp[S_{pert}^{NLL}] \to 1+S_{pert}^{FXO}$ 
turns out to be unreliable on a wide portion of the $b_T$ space, so that the required condition $W^{FXO} \sim W^{NLL}$ at
$q_T\sim Q$ is not fulfilled.

We noticed also that, at small  $b_T$, the Sudakov factor does not converge to zero, 
as it should~\cite{Altarelli:1984pt, Frixione:1998dw}. We tested one of the available prescriptions 
to correct for this unphysical behaviour, Eq.~\eqref{w-prescr}, and we found that, for intermediate $Q^2$ values, 
the region of $b_T$  modified by this correction is large enough to have an impact on the Sudakov factor, 
while at higher $Q^2$ its impact is totally negligible. 
Using all these recipes we find that, at intermediate HERA-like energies, the  $b_T$ variation of $S_{pert}$ is limited, 
finally allowing for a successful expansion $\exp[S_{pert}^{NLL}] \to 1+S_{pert}^{FXO}$. 
Consequently, we found a region in the $q_T$ space where $W^{FXO} \sim W^{NLL}$: here a matching could be attempted. 

However, the matching procedure of Eq.~\eqref{match2} 
is still affected by a number of difficulties.
First of all, the condition $W^{FXO} \sim W^{NLL}$ is fulfilled when $q_T$ is larger than $Q$, rather than $q_T\sim Q$ 
as one would have expected.
Secondly, this procedure requires a second point of matching, at low $q_T$, where one should switch to $W^{NLL}$. 
One can choose (as we did) the point in which $W^{FXO} = d\sigma^{NLO}$, but this choice is totally arbitrary and is not 
supported by any physical motivation.
Therefore, one can well wonder whether a direct switch from $W^{NLL}$ to $d\sigma^{NLO}$ at smaller values of $q_T$ 
could not be more appropriate~\cite{Arnold:1990yk}. 
Fig.~\ref{f11} shows that this direct switch 
is actually possible at $q_T \sim 15$ GeV. 
This prescription is as unpredictive as the previous one, but indeed easier to implement.

Not surprisingly, the resummation scheme in $b_T$ space with the $b_*$ prescription, although successful in some kinematical 
configurations, has proven to be quite controversial and of difficult implementation, when it is stretched to the region of 
low $Q^2$ and/or large $q_T$.
Therefore, other theoretical and phenomenological studies are required in order to find the appropriate description for these regions.

Indeed, being the non-perturbative details of such importance to the description of the cross section, 
the extension of our work to other methods applied in the literature to treat the non-perturbative 
part~\cite{Ellis:1997ii,Kulesza:1999gm,Qiu:2000hf,Bozzi:2005wk,Koike:2006fn}, 
deserves further studies.

We emphasize the importance of
having experimental data available in order to test all the mechanisms developed in soft gluon resummation and
study the non-perturbative aspects of the nucleon.
It is essential to have (and analyze) data from HERA($\sqrt{s}=300$ GeV), Electron-Ion Collider
($\sqrt{s}=20$ -- $100$ GeV), COMPASS ($\sqrt{s}=17$ GeV), HERMES ($\sqrt{s}=7$ GeV), and Jefferson Lab 12 ($\sqrt{s}=5$ GeV).
In particular, it will be very important to study experimental data on $q_T$ distributions that span the region of low $q_T\ll Q$
up to the region of $q_T\sim Q$.

\bigskip

\section*{Acknowledgements}
We thank M. Anselmino, J. Collins, J. Qiu, Z. Kang, P. Sun and F. Yuan for useful discussions. 
A.P. acknowledges support by the U.S. Department of Energy, Office of Science, Office of Nuclear Physics, under contract No. DE-AC05-06OR23177.
  M.B. and S.M. acknowledge support from the European Community under the FP7
``Capacities - Research Infrastructure'' program (HadronPhysics3, Grant Agreement 
283286),  
and support from the ``Progetto di Ricerca Ateneo/CS'' (TO-Call3-2012-0103). 

\appendix
\section{Fixed order cross section \label{A}}
The NLO FXO cross section for SIDIS processes is obtained from Eq.~(\ref{eq:W}) with 
the resummed $W$-term, expanded at first order in $\alpha_s$, written in the following form   
\bea
W^{FXO}(x,z,b_T,Q)\!&=&\!
\sum_q\! e_q^2 \Bigg\{ \Big(1+S^{(1)}-4C_F \frac{\alpha_s(\mu_b)}{\pi}\Big) \, f_q(x,\mu_b^2)\, D_q(z,\mu_b^2)\nonumber \\
&+& \frac{\alpha_s(\mu_b)}{2\pi} 
\left( f_q(x,\mu_b^2) 
\left[C_F\!\!\int_z^1 \! \frac{dz^\prime}{z^\prime}\Big((1-z^\prime) + 
2\ln z^\prime \frac{1+{z^\prime}^2}{1-z^\prime}\Big) \, D_q(z/z^\prime,\mu_b^2)  \right.\right. \nonumber \\
&+& \left.\left. \Big(z^\prime + 2\ln z^\prime \frac{1+(1-z^\prime)^2}{z^\prime}\Big) \, D_g(z/z^\prime,\mu_b^2)\right] 
\right.\nonumber \\
&+& \left.
D_q(z,\mu_b^2) \left[ \int_x^1 \frac{dx^\prime}{x^\prime}\Big(C_F(1-x^\prime) f_q(x/x^\prime,\mu_b^2) 
\right. \right. \nonumber \\
&+& \left.\left. T_F\, x^\prime (1-x^\prime) f_g(x/x^\prime,\mu_b^2)
\right)
\bigg]\Bigg)
\right\}\, , \label{eq:FXO}
\eea
where $S^{(1)}$  is the NLL Sudakov form  factor 
\be
S^{(1)}=- \int_{\mu_b^2}^{Q^2} \frac{d\mu^2}{\mu^2}\frac{\alpha_s(\mu)}{\pi}\Big(A^{(1)}\ln\left(\frac{Q^2}{\mu^2}\right)+B^{(1)}\Big)\,.
\ee

\section{Correspondence between CSS resumation and TMD evolution at first order in the strong coupling\label{sec:TMD}}

The CSS resummation of Ref.~\cite{Collins:1984kg} and the Collins TMD evolution scheme~\cite{Collins:2011zzd} are closely related. 
An obvious advantage of the scheme of Ref.~\cite{Collins:2011zzd} is that both TMD PDF and TMD FF are well defined operators,
while the original Ref.~\cite{Collins:1984kg}  deals with the whole cross-section.

In this appendix we will briefly outline how the CSS main formula for the SIDIS cross section, Eq.~\eqref{SIDIS-CSS1}, 
can be derived from the TMD evolution framework presented in Ref.~\cite{Collins:2011zzd}.
Using TMD factorization the unpolarized SIDIS cross section can be written as:
\begin{equation}
\frac{d\sigma}{ dx\,dy \,dz \,d q_T^2}=\pi z^2 H^2(Q;\mu)\int\frac{d^2 \boldsymbol{b}_T e^{i \boldsymbol{q}_T\cdot\boldsymbol{b_T}}}{(2\pi)^{2}}\Bigg\{
\sum_{j}e^2_j\tilde{F}_{j}(x,b_T,\mu,\zeta_{F})\tilde{{D}}_{j}(z,b_T,\mu,\zeta_{D})\Bigg\} + Y\, ,\label{Master_RAC1}
\end{equation}
where $H^2(Q;\mu)$ is a process dependent hard factor~\cite{Collins:2011zzd, Aybat:2011vb}. Setting $\mu=Q$, we obtain:
\begin{equation}
 H^2(Q;Q)=\sigma_{0}^{DIS}\left\{1-\frac{\alpha_s(Q)}{\pi} (-4 C_F) +{\cal O} (\alpha_s^2)\right\}\,.\label{eq:hs}
\end{equation}
The TMD PDF  $\tilde{F}_{q}(x,b_T,Q,\zeta_F)$ is given by
\begin{eqnarray}
\tilde{F}_{j}(x,b_T,Q,\zeta_{F})&=&\left({\frac{\sqrt{\zeta_F}}{\mu_b}}\right)^{\tilde{K}(b_*,\mu_b)}
\sum_j\int_x^1 \frac{d \hat x}{\hat x} \tilde{C}_{ji}^{in}(x/\hat x, b_*,\mu_b,\mu_b^2) f_i(\hat x,\mu_b)\nonumber\\
&&\times\exp\left\{\int_{\mu_b}^Q \frac{d \mup}{\mup} \left(\gamma_F(\mup;1)-\ln\left(\frac{\sqrt{\zeta_F}}{\mup}\right)\gamma_K(\mup)\right)\right\}\nonumber\\
&&\times \exp\left\{-g_{P}(x,b_T)-g_K(b_T)\ln\left(\frac{\sqrt{\zeta_F}}{\sqrt{\zeta_{F0}}}\right)\right\}\label{Master_RAC4}\,,
\end{eqnarray}
similary, the TMD FF is 
\begin{eqnarray}
\tilde{D}_{j}(z,b_T,Q,\zeta_{D})&=&\left({\frac{\sqrt{\zeta_D}}{\mu_b}}\right)^{\tilde{K}(b_*,\mu_b)}
\sum_k\int_z^1 \frac{d \hat z}{\hat z^3}\tilde{C}_{kj}^{out}(z/\hat z, b_*,\mu_b,\mu_b^2) D_j(\hat z,\mu_b)\nonumber\\
&&\times\exp\left\{\int_{\mu_b}^Q \frac{d \mup}{\mup} \left( \gamma_D(\mup;1)-\ln\left(\frac{\sqrt{\zeta_D}}{\mup}\right)\gamma_K(\mup)\right) \right\}\nonumber\\
&&\times\exp\left\{-g_{H}(z,b_T)-g_K(b_T)\ln\left(\frac{\sqrt{\zeta_D}}{\sqrt{\zeta_{D0}}}\right)\right\}\label{Master_RAC4_FF}\,.
\end{eqnarray}
Here $g_P(x,b_T)$, $g_{H}(z,b_T)$ and $g_K(b_T)$ are non-perturbative functions that correspond to intrinsic quark motion in the proton and the final hadron and to the universal function that describes non perturbative behaviour of soft gluon radiation.
Rapidity divergence regulators, as explained in Ref.~\cite{Collins:2011zzd}, $\zeta_F$ and $\zeta_D$ appear in the TMD PDF and FF to obtain a well defined operator definition.
These regulators are such that $\zeta_F \zeta_D\approx Q^4$. In principle the cross section of Eq.~\eqref{Master_RAC1} is independent of $\zeta_F$ and $\zeta_D$,
therefore one can conveniently choose $\zeta_F=\zeta_D\equiv Q^2$, and similarly  $\zeta_{F0}=\zeta_{D0}=Q^2_0$.

The kernel $\tilde{K}$ encodes the $\zeta$ dependence of TMDs,  $\gamma_K$ is the so-called cusp anomalous dimension \cite{Korchemsky:1987wg} while $\gamma_F$, $\gamma_D$ are the anomalous dimensions of 
$\tilde F$, $\tilde D$.
We will use the first loop expressions of $\tilde{K}$, $\gamma_K$ and $\gamma_F$ from Refs.~\cite{Collins:2011zzd,Aybat:2011zv}
\begin{eqnarray}
\tilde{K}(b_T,\mu)&=&-\frac{\alpha_s C_F}{\pi}\ln\left(\frac{\mu^2 b_T^2}{C_1^2}\right)\, ,\nonumber\\
\gamma_K(\mu)&=&2\frac{\alpha_s(\mu) C_F}{\pi}\nonumber\, ,\\
\gamma_D(\mu,\zeta/\mu^2)=\gamma_F(\mu,\zeta/\mu^2)&=&\frac{\alpha_s(\mu) C_F}{\pi}\left(\frac{3}{2}-\ln\left(\frac{\zeta}{\mu^2}\right)\right)\,,
\label{eq:oneloop}
\end{eqnarray}
and perform our comparison with CSS at one loop as well.
One can easily check that:
\begin{eqnarray}
 \tilde{K}(b_*,\mu_b)&\equiv& 0\label{eq:KTilde}\\
 \int_{\mu_b}^Q \frac{d \mup}{\mup}  \left(\gamma_F(\mup;1)-\gamma_K(\mup)\ln\left(\frac{\sqrt{\zeta}}{\mup}\right)\right)
 &=& \int_{\mu_b}^Q \frac{d \mup}{\mup} \gamma_F(\mu;\zeta/\mu^2)\label{eq:gf2}\,.
\end{eqnarray}
Since  $\zeta_F=\zeta_D\equiv Q^2$, we have:
\begin{eqnarray}
 \int_{\mu_b}^Q \frac{d \mup}{\mup}\gamma_F(\mu;Q^2/\mu^2)&=&\int_{\mu_b}^Q \frac{d \mup}{\mup}\frac{\alpha_s(\mu) C_F}{\pi}\left(\frac{3}{2}-\ln\left(\frac{Q^2}{\mu^2}\right)\right)\nonumber\\
&=&-\frac{1}{2}\int_{\mu_b^2}^{Q^2} \frac{d\mu^2}{\mu^2}\frac{\alpha_s(\mu)}{\pi}\Big(A^{(1)}\ln\left(\frac{Q^2}{\mu^2}\right)+B^{(1)}\Big)\nonumber\\
&=&\frac{1}{2} S_{pert}(b_*,Q)\,,\label{eq:gammafS}
\end{eqnarray}
where $S_{pert}(b_T,Q)$ is the same perturbative Sudakov factor defined in the CSS scheme, Eq.~\eqref{S}, calculated at first order in $\alpha_s$. 
The same expression holds for the integral of $\gamma_D$.

The Wilson coefficients in Eqs.~\eqref{Master_RAC4} and \eqref{Master_RAC4_FF} are always evaluated at the scales $\mu=\mu_b$ and $\zeta=\mu_b^2$,
therefore their expressions simplify considerably:
\begin{eqnarray}
\tilde{C}_{qq'}^{\rm(0)in}(x,b_*,\mu_b,\mu_b^2)&=&\delta_{qq'} \delta(1-x)\equiv C_{qq'}^{\rm(0)in}(x) \label{eq:firstCTMDCSS}\\
\tilde{C}_{qq'}^{\rm (0)out}(z,b_*,\mu_b,\mu_b^2)&=&\delta_{qq'} \delta(1-z) \equiv {C}_{qq'}^{\rm (0)out}(z)\\
\tilde{C}_{gq}^{\rm(0)out}(z,b_*,\mu_b,\mu_b^2) &=& 0 \equiv {C}_{gq}^{\rm(0)out}(z) \\
\tilde{C}_{qg}^{\rm(0)in}(x,b_*,\mu_b,\mu_b^2)&=&0\equiv C_{qg}^{\rm(0)in}(x)
\end{eqnarray}
and
\begin{eqnarray}
\tilde{C}_{qq'}^{\rm(1)in}(x,b_*,\mu_b,\mu_b^2)&=& \delta_{qq'}\frac{C_F}{2} \Big\{(1-x) \Big\}\equiv C_{qq'}^{\rm(1)in}(x)+\delta_{qq'} 2 C_F \delta(1-x)\\
\tilde{C}_{qg}^{\rm(1)in}(x,b_*,\mu_b,\mu_b^2)&=&T_F[x(1-x)]\equiv C_{qg}^{\rm(1)in}(x) \\
\tilde{C}_{qq'}^{\rm (1)out}(z,b_*,\mu_b,\mu_b^2)&=&\delta_{qq'}\frac{C_F}{2 z^2}\left\{(1-z)+2 \ln(z)\left[\frac{1+z^2}{1-z}\right]\right\}\nonumber\\
&\equiv&\frac{1}{z^2}\tilde{C}_{qq'}^{\rm (1)out}(z)+\frac{1}{z^2}\delta_{qq'} 2 C_F \delta(1-z)\\
\tilde{C}_{gq}^{\rm(1)out}(z,b_*,\mu_b,\mu_b^2)&=&\frac{C_F}{2 z^2} \left\{z+2 \ln(z)\frac{1+(1-z)^2}{z}\right\}\equiv \frac{1}{z^2}{C}_{gq}^{\rm(1)out}(z)\,. \label{eq:lastCTMDCSS}
\end{eqnarray}
By defining 
\begin{equation}
S_{NP} =-g_P(x,b_T)- g_{H}(z,b_T) -2 g_K(b_T)\ln\left(\frac{Q}{Q_0}\right)\,,
\end{equation}
and making use of Eqs.~\eqref{eq:hs}-\eqref{Master_RAC4_FF} and \eqref{eq:KTilde}-\eqref{eq:gammafS} we can rewrite Eq.~\eqref{Master_RAC1} as:
\begin{equation}
\frac{d\sigma}{ dx\,dy \,dz \,d q_T^2}=\pi\sigma_{0}^{DIS}\int\limits_{0}^{\infty}
\frac{d {b}_T b_T}{(2\pi)} J_0( {q}_T {b_T})
W^{TMD}(x,z,b_*,Q)\exp\left[S_{N\!P}(x,z,b_T,Q)\right]+Y \, , \label{eq:TMD2}
\end{equation}
where
\begin{eqnarray}
W^{TMD}(x,z,b_*,Q)&=& \left\{1-\frac{\alpha_s(Q)}{\pi} (-4 C_F) \right\} \exp\left[S_{pert}(b_T,Q)\right]\nonumber\\
&&\times \sum_j\! e_j^2\sum_{i,k}
\left[\tilde{C}_{ji}^{\rm in}\otimes f_{i}(x,\mu_b^2)\right]\,
\left[\left(\tilde{C}_{kj}^{\rm out} z^2\right)
\otimes D_{k}(z,\mu_b^2)\right]\label{eq:TMDCSS1}\,.
\end{eqnarray}
The symbol $\otimes$ stands for the usual convolution defined in Eq.~\eqref{eq:convolution}.
Notice that in Eq.~\ref{eq:TMDCSS1} we use the identity:
\begin{eqnarray}
\int_z^1 \frac{d \hat z}{\hat z^3}\tilde{C}^{out}(z/\hat z) D(\hat z)&=&  \int_z^1 \frac{d \hat z}{ \hat z} C(z/\hat z) \frac{D(\hat z)}{\hat z^2} = \left[\tilde{C}^{out} \otimes \frac{D(z)}{z^2}\right] \nonumber\\
&=& \frac{1}{z^2} \int_z^1 \frac{d \hat z}{ \hat z} \frac{z^2}{\hat z^2} C(z/\hat z) D(\hat z) = \frac{1}{z^2} \left[(\tilde{C}^{out} z^2)\otimes D(z)\right]\,.
\end{eqnarray}
Finally substituting Eqs.~\eqref{eq:firstCTMDCSS}-\eqref{eq:lastCTMDCSS} in Eq.~\eqref{eq:TMDCSS1}, and neglecting terms of order $\alpha_s^2$ in the product between
the convolutions and the hard factor $H$, we have 
\begin{eqnarray}
W^{TMD}(x,z,b_*,Q)&\simeq& \exp\left[S_{pert}(b_T,Q)\right]\nonumber\\
&&\times \sum_j\! e_j^2\sum_{i,k}
\,{C}_{ji}^{\rm in}\otimes f_{i}(x,\mu_b^2)\; {C}_{kj}^{\rm out}
\otimes D_{k}(z,\mu_b^2) + {\cal O} (\alpha_s^2)\,,
\label{eq:fWTMD}
\end{eqnarray}
which corresponds to the resummed cross section $W^{SIDIS}$ of Eq.~\eqref{eq:W}, calculated up to first order in $\alpha_s$ in the Wilson coefficients and the Sudakov form factor.
Therefore, the difference between the TMD formalism of Ref.~\cite{Collins:2011zzd} and the original CSS scheme of Ref.~\cite{Collins:1984kg}
is of higher order in perturbative theory.

\bibliographystyle{JHEP}
\bibliography{\BibPath/sample}

\end{document}